\documentclass[11pt,a4paper]{article}
\usepackage{jcappub}
\usepackage{graphicx}
\usepackage{longtable}
\usepackage{amsmath}
\usepackage{amssymb}
\usepackage{subfigure}
\usepackage{hyperref}
\usepackage{dcolumn}
\usepackage{multirow}
\usepackage{bm}
\usepackage{color}

\newcommand{\be}{\begin{eqnarray}}
\newcommand{\ee}{\end{eqnarray}}
\newcommand{\mpl}{{M_{\rm {pl}}}}

\newcommand{\dd}{\, {\rm d}}
\newcommand{\gsim}{\;\mbox{\raisebox{-0.5ex}{$\stackrel{>}{\scriptstyle{\sim}}$}
}\;}
\newcommand{\lsim}{\;\mbox{\raisebox{-0.5ex}{$\stackrel{<}{\scriptstyle{\sim}}$}
}\;}
\def\eea{\end{eqnarray}}
\def\bea{\begin{eqnarray}}

\newcommand{\pn}{\Phi_{\rm N}}

\newcommand{\tg}{\tilde{g}}
\newcommand{\iii}{_{\rm i}}
\newcommand{\eff}{_{\rm eff}}

\newcommand{\nm}{{\mu\nu}}
\newcommand{\pmi}{\phi_{\rm min}}
\newcommand{\mmm}{{_{\rm m}}}
\newcommand{\rmm}{{\rho_{\rm m}}}
\newcommand{\vpp}{\varphi}
\newcommand{\inff}{_\infty}
\newcommand{\phii}{\phi_\infty}
\allowdisplaybreaks
\begin{document}
\title{Disformal Theories of Gravity: From the Solar System to Cosmology}
\author[a,b]{Jeremy Sakstein}
\affiliation[a]{DAMTP, Centre for Mathematical Sciences, University of Cambridge, Wilberforce Road, Cambridge CB3 0WA, UK}
\affiliation[b]{Perimeter Institute for Theoretical Physics, 31 Caroline St. N,
Waterloo, ON, N2L 6B9, Canada}
\emailAdd{j.a.sakstein@damtp.cam.ac.uk}

\abstract{This paper is concerned with theories of gravity that contain a scalar coupled both
conformally and disformally to matter through the metric. By systematically deriving the non-relativistic limit, it is shown that
no new non-linear screening mechanisms are present beyond the Vainshtein mechanism and chameleon-like screening. If
one includes the cosmological expansion of the universe, disformal effects that are usually taken to be absent can be present in
the solar system. When the conformal factor is absent, fifth-forces can be screened on all scales when the
cosmological field is slowly-rolling. We investigate the cosmology of these models and use local tests of gravity to place new
constraints on the disformal coupling and find $\mathcal{M}\gsim\mathcal{O}(\textrm{eV})$, which is not competitive with
laboratory tests.
Finally, we discuss the future prospects for testing these theories and the implications for other theories of modified gravity.
In particular, the Vainshtein radius of solar system objects can be altered from the static prediction when cosmological
time-derivatives are non-negligible.}
\maketitle

\section{Introduction}

Modified theories of gravity (see \cite{Clifton:2011jh} for a review) have received a renewed interest over the last decade for a
variety of reasons, most notably as an alternative to exotic matter such as quintessence to explain the acceleration of the
cosmic expansions. Among them, theories that contain \textit{screening mechanisms} (see \cite{Jain:2010ka,Joyce:2014kja} for a
review)
have been particularly well-studied due to their ability to hide any effects of the modifications on small scales using non-linear
effects. In particular, the chameleon mechanism \cite{Khoury:2003aq,Khoury:2003rn}, the symmetron effect
\cite{Hinterbichler:2010es} and the environment-dependent Damour-Polyakov effect \cite{Brax:2010gi} all screen by suppressing the
local scalar charge-to-mass ratio whilst the Vainshtein mechanism \cite{Vainshtein:1972sx} screens in galileon
\cite{Nicolis:2008in} and massive gravity theories \cite{Hinterbichler:2011tt,deRham:2014zqa} by suppressing scalar gradients. 

These mechanisms all arise in theories whose covariantisations include a conformal coupling to matter via the metric
($\tg_\nm=A^2(\phi)g_\nm$) and so one may wonder whether an extension to \textit{disformal} couplings of the form
$B^2(\phi)\partial_\mu\phi\partial_\nu\phi$---these were first studied by \cite{1992mgm..conf..905B,Bekenstein:1992pj}---may hold interesting
properties\footnote{One can extend these theories to include functions of the kinetic term but we will not consider those here.
Interestingly, these theories can give rise to second order equations of motion despite appearing to be higher-derivative at the
level of the action \cite{Zumalacarregui:2013pma}.}. Indeed, these have been studied in the context of
quintessence \cite{Koivisto:2008ak,Zumalacarregui:2010wj} but recently they have begun to be studied in the context of dark energy and modified
gravity i.e. the scalar has been allowed to couple directly to matter\footnote{See \cite{Kaloper:2003yf} for an inflationary scenario using this 
coupling and the resultant bounds from particle physics 
constraints.} \cite{Koivisto:2012za,Zumalacarregui:2012us,vandeBruck:2012vq,Brax:2013nsa,vandeBruck:2013yxa,Brax:2014vva,Brax:2014zba}. In
terms of local screening, \cite{Zumalacarregui:2012us,vandeBruck:2012vq} have studied the behaviour of these theories on a
Minkowski background and have asserted the existence of a so-called \textit{disformal screening mechanism} whereby scalar
gradients are suppressed and time-derivatives evolve towards zero in dense environments. Since then, there has been no work
towards
classifying the conditions under which this is realised and no attempts to constrain the model parameters locally. This is an
important step towards looking for potential astrophysical signatures.

In this work, we re-examine the non-relativistic limit of these theories, not by taking the space-time to be Minkowski but instead
by performing a systematic expansion in the metric potentials sourced by the presence of a non-relativistic object. We do this
for a metric containing both conformal and disformal couplings. We find that when the space-time being perturbed is Minkowski,
the non-relativistic limit is identical to a scalar-tensor theory that contains a conformal coupling only and hence there are no
new screening mechanisms beyond those mentioned above. In particular, this means that no screening mechanism is required in the absence of conformal 
couplings. This is what was found by \cite{Noller:2012sv}. The appearance of the
disformal screening mechanism arises when one prioritises post-Newtonian corrections over the leading-order Newtonian ones and
makes the further assumption that one can neglect the radial dependence of the field.

When instead one considers perturbations about a Friedmann-Robertson-Walker (FRW)
space-time---which is the physically realistic scenario---one finds a correction to the local scalar charge that depends on a
combination of the disformal model parameters and the cosmological time-derivatives of the homogeneous component of the field. It
is zero whenever the time-derivatives vanish but non-zero when this is not the case, even when the conformal factor is absent. In this case, a 
screening mechanism is mandated in order to avoid order-one fifth-forces on all scales. Whereas there may be specific models where the cosmological 
dynamics screen both the conformal and disformal contribution to the field's source, we specialise to the case of a disformal coupling only, which we 
argue can be
screened generically in any quintessence-like model. This is because we expect the field to slowly roll down its potential and
hence have small time-derivatives and a negligible scalar charge at late times.

In order to study this quantitatively, we introduce a simple model where a disformally coupled scalar rolls down a quadratic
potential. We integrate the Friedmann-scalar field equations numerically and find that a large disformal coupling can delay the
time at which the scalar field begins to roll (i.e. the onset of dark energy domination). Furthermore, we show how the local
scalar charge is pushed to zero by the cosmological evolution towards the minimum.

Next, we use local tests of gravity to constrain the model parameters. The theory obeys the equivalence principle, which
restricts the number of tests we can use compared with chameleon models. Despite this, we are able to use the Cassini bound on
the parametrised-post-Newtonian (PPN) parameter $\gamma_{\rm PPN}$ to place the strongest constraint. Using the Lunar Laser
Ranging bound on the time-variation of Newton's constant we obtain a weaker bound. Laboratory tests cannot probe these
theories because the local scalar charge is zero unless the field's mass is of order Hubble\footnote{We discuss some possible
exceptions to this.}. Comparing our results with those of \cite{Kaloper:2003yf,Brax:2014vva,Brax:2014zba}, who study the quantum loop
corrections arising from the disformal coupling $T\mmm^\nm\partial_\mu\phi\partial_\nu\phi/\mathcal{M}^4$, we find that
$\mathcal{M}\gsim\mathcal{O}(\textrm{eV})$ whereas they find $\mathcal{M}\gsim\mathcal{O}(10^2\textrm{GeV})$ using mono-photon
searches at the LHC. Our constraints are then not competitive.

Finally, we discuss the implications of the results presented here. First, we find that when our bound is imposed, the cosmology
of our model is indistinguishable from the equivalent quintessence model, although it remains to be seen whether this is a
generic feature of disformally coupled theories. Having elucidated exactly how these theories screen, we discuss the future
prospects for improving the constraints and identify laboratory tests along the lines of \cite{Kaloper:2003yf,Brax:2014vva,Brax:2014zba} and
linear and non-linear cosmological probes as promising avenues. Other theories of modified gravity that include a disformal
coupling often neglect it on small scales. We end by discussing how one should include a time-dependent vacuum expectation value
(VEV) corresponding to the cosmological value of the field. If this is omitted then potentially large fifth-forces that are
predicted by the theory are absent from the calculation. In the context of Vainshtein screened theories, the disformal coupling
can change the effective matter coupling---usually taken to be $\mathcal{O}(1)$---and alter the Vainshtein radius.

This paper is organised as follows: In section \ref{sec:2} we introduce disformal theories of gravity, define the model
parameters and set out our conventions. These differ from several other works in the literature but have the
advantage of making the local behaviour of the field clearer and give us dimensionless quantities with which to construct the
non-relativistic limit. This limit is systematically constructed in section \ref{sec:nrlim} where the screening properties of the
theory are discussed. Next, we specialise to a simple model that we expect to realise screening at late
times and study its cosmology in section \ref{sec:cosmo}. In section \ref{sec:tests} we place new bounds on the model parameters
using local tests of gravity and compare them with other recent bounds. Finally, the
implications of the results presented here are discussed in section \ref{sec:disc} before concluding in section \ref{sec:concs}.

\section{Disformal Theories of Gravity}
\label{sec:2}
Disformal theories of gravity are described by the following action:
\begin{equation}\label{eq:act}
 S=\mpl^2\int\dd^4x\sqrt{-g}\left[\frac{R}{2}+X-V(\phi)\right]+S_{\rm
m}\left[\tg;\Psi\iii\right],
\end{equation}
where the various matter fields $\Psi\iii$ are coupled to the Jordan frame metric
\begin{equation}\label{eq:metric}
 \tg_{\mu\nu}=A^2(\phi)\left(g_{\mu\nu}+\frac{B^2(\phi)}{\Lambda^2}\partial_\mu\phi\partial_\nu\phi\right),
\end{equation}
and $X\equiv -1/2\nabla_\mu\phi\nabla^\mu\phi$. We will refer to $A$ and $B$ as the \textit{conformal} and \textit{disformal}
factors respectively\footnote{Note that our definition of $\tg_{\mu\nu}$ differs from
\cite{Koivisto:2012za,Zumalacarregui:2012us}. Our definition has the advantage that it does not mix conformal and disformal
effects in the equations of motion. Note also that we are using a dimensionless field, which helps to make the connection with the
Newtonian limit of general relativity clearer. Sending $\phi\rightarrow\phi/\mpl$ in (\ref{eq:metric}), the two conventions are
related via the transformations $B^2A^2/\mpl^2\Lambda^2\rightarrow B(\phi)$ and $A^2\rightarrow A(\phi)$.}. Furthermore, we define
the quantities
\begin{equation}\label{eq:agdefs}
 \alpha(\phi)\equiv\frac{\dd\ln A(\phi)}{\dd \phi}\quad \textrm{and}\quad\gamma(\phi)\equiv\frac{\dd\ln B(\phi)}{\dd \phi}.
\end{equation}
Since the field is coupled directly to the matter, the energy momentum tensor $T\mmm^\nm\equiv 2/\sqrt{-g}\delta S\mmm/\delta
g_{\mu\nu}$ is not covariantly conserved and instead one has
\begin{equation}\label{eqtmunucons}
 \nabla_\mu T^{\mu\nu}=-Q\nabla^\nu\phi,
\end{equation}
where
\begin{equation}\label{eq:Qtmunudef}
 Q\equiv
\nabla_\mu\left(\frac{B^2(\phi)}{\Lambda^2}T\mmm^{\mu\nu}\nabla_\mu\phi\right)-\alpha(\phi)T\mmm-\frac{B(\phi)^2}{\Lambda^2}
\left[\gamma(\phi)+\alpha(\phi)\right]T\mmm^{\mu\nu}\nabla_\mu\phi\nabla_\nu\phi,
\end{equation}
where $T\mmm\equiv g_{\mu\nu}T\mmm^{\mu\nu}$ is the trace of the energy-momentum tensor. The equation of motion for the scalar is 
\begin{equation}
 \Box\phi=V(\phi)_{,\,\phi}+Q,
\end{equation}
although this is not particularly useful because $Q$ contains derivatives of $T\mmm^{\mu\nu}$. Contracting equation
(\ref{eqtmunucons}) with $\nabla_\nu\phi$, one can obtain an expression for $\nabla_\mu T\mmm^{\mu\nu}\nabla_\nu\phi$, which may
be substituted back into equation (\ref{eq:Qtmunudef}) to find
\begin{align}
\chi\Box\phi&-8\pi G\frac{B^2}{\Lambda^2}T_{\rm m}^{\mu\nu}\nabla_\mu\nabla_\nu\phi=-8\pi\alpha
GT_{\rm m}-8\pi G\frac{B^2}{\Lambda^2}\left(\alpha-\gamma\right)T_{\rm
m}^{\mu\nu}\partial_\mu\phi\partial_\nu\phi+\chi V(\phi)_{,\phi}.\label{eq:sfeom},
\end{align}
where
\begin{equation}
 \chi\equiv 1-\frac{2B^2X}{\Lambda^2}
\end{equation}
We will work with this form of the equation of motion exclusively. The Einstein equations are
\begin{equation}\label{eq:EE}
 G^\mu_\nu=8\pi G\left(T\mmm_{\,\mu}^{\nu}+T_{\phi\,\mu}^{\nu}\right),
\end{equation}
where 
\begin{equation}\label{eq:tmnphi}
 T_{\phi\,\mu\nu}=\frac{1}{8\pi
G}\left[\nabla_\mu\phi\nabla_\nu\phi-\delta_\mu^\nu\left(\frac{1}{2}\nabla_\mu\phi\nabla^\mu\phi+V(\phi)\right)\right]
\end{equation}
is the energy-momentum tensor of the field. Again, this is not conserved and one has
$\nabla_{\mu}T_\phi^{\mu\nu}=-\nabla_\mu T\mmm^{\mu\nu}=Q\nabla^\nu\phi$.

\section{The Non-Relativistic Limit and Screening}
\label{sec:nrlim}
We are interested in the behaviour of these theories in the solar system and so we must examine the non-relativistic limit of both
the field and geodesic equations in the presence of a non-relativistic source. In order to simplify the discussion
and to elucidate the physics, we will first derive the non-relativistic limit for a configuration that is an inhomogeneous (but
not linear) perturbation about Minkowski space. We will see that the non-relativistic limit is identical to that of
any scalar-tensor theory with a scalar potential and conformal coupling only and is therefore unscreened unless one
invokes chameleon or other similar mechanisms. Next, we will present the equivalent results in an FRW universe, leaving the
details in appendix \ref{sec:FRWnewt}. In this case, we will find an additional contribution to the equations of motion coming
from the
time-variation of the homogeneous component of the field. 

\subsection{Minkowski Space}

\subsubsection{Non-Relativistic Limit of the Field Equations} 

One may na\"ively assume that the non-relativistic limit of equation (\ref{eq:sfeom}) corresponds to taking
$g_{\mu\nu}=\eta_{\mu\nu}$ and, indeed, in many scalar-tensor theories this is sufficient. In disformal theories it is
not. Minkowski space is a vacuum solution of the Einstein equations (\ref{eq:EE}) i.e. when $T\mmm^{\mu\nu}=T_\phi^\nm=0$.
In particular, this means that the homogeneous component of the scalar can at most be a constant everywhere in space and
time\footnote{For the purposes of this discussion, the scalar potential is assumed to have a minimum at $\pmi$ such that
$V(\pmi)=0$. If this is not the case then the appropriate solutions are de-Sitter or anti-de-Sitter space depending on the sign
of $V$. We will relax this
assumption when we discuss the case of an FRW background.}. We now wish to introduce a non-relativistic source with
energy-momentum tensor $T_{\mu\nu}=\textrm{diag}(\rho,0,0,0)$, which will source a perturbation to this background described by
the metric potentials $\Phi$ and $\Psi$ in the conformal Newtonian gauge:
\begin{equation}\label{eq:minkpert}
 \dd s^2=(-1-2\Phi)\dd t^2+(1-2\Psi)\dd\vec{x}^2.
\end{equation}
The appropriate equation for the inhomogeneous component of the field is then equation (\ref{eq:sfeom}) with $g_\nm$
corresponding to this perturbed line element. Since the inhomogeneous component of the field is sourced by the perturbation, it
must be at least first-order in the metric potentials, which is why it is acceptable to ignore cross terms of the schematic form
$\Psi\phi$ etc. but this is not to say that the resultant equation does not contain any other post-Newtonian terms. In short, the
non-relativistic limit corresponds to an expansion in the metric perturbations and the field about the vacuum solution of the
system whereas taking $g_{\mu\nu}=\eta_{\mu\nu}$ yields and equation for a scalar field on a fixed Minkowski background.
In order to determine how each quantity in (\ref{eq:sfeom}) compares with the metric potential, one must study the linearised
Einstein equations, which are
\begin{align}
\nabla^2\Psi&=4\pi
G\rho\mmm+\frac{1}{4}\partial_0\phi\partial^0\phi+\frac{1}{4}\partial_i\phi\partial^i\phi+\frac{1}{2}V(\phi)\label{eq:poisslin},\\
\partial_0\partial_i\Psi&=-\frac{1}{2}\partial_0\phi\partial_i\phi,\label{eq:0i}\\
\left(\partial_i\partial_j-\frac{1}{3}\delta_{ij}
\nabla^2\right)\left(\Psi-\Phi\right)&=\partial_i\phi\partial_j\phi-\frac { 1 } { 3 }
\delta_{ij}\partial_k\phi\partial^k\phi\quad\textrm{and}\\
\partial_0\partial_0\Psi+\frac{1}{3}\nabla^2\left(\Phi-\Psi\right)&=-\frac{1}{6}\left(\frac{3}{2}
\partial_0\phi\partial_0\phi+\frac{1}{2}\partial_k\phi\partial^k\phi-3V\right)\label{eq:ij}.
\end{align}
Since we are working in the Einstein frame, the equations of motion for $g_{\mu\nu}$ are identical to those of general
relativity sourced by the energy-momentum tensor for both matter and the scalar field. Before worrying about modified
gravity effects, we must ensure that the correct Newtonian limit of general relativity is recovered i.e. that the scalar does not
source any deviations of the Newtonian potential $\pn$ from general relativity\footnote{This is an Einstein frame statement,
where the system behaves as general relativity and a fifth-force due to the scalar. In the Jordan frame, one will
instead find a modification of the Poisson equation linear in $\nabla^2\phi$.}. The Newtonian limit corresponds to
ignoring time-derivatives of the metric potentials to find $\nabla^2\Phi=4\pi G\rho\mmm$ and $\Phi=\Psi$. In this case, $\Phi$
is identified with the Newtonian potential $\pn$ and we will work with this from here on. Any corrections to equations
(\ref{eq:poisslin}) to (\ref{eq:ij}) coming from the scalar
must then be post-Newtonian and we must therefore impose that $\partial_0\phi\partial^0\phi$, $\partial_i\phi\partial^i\phi$ and
$\partial_0\phi\partial_i\phi$ are of order $\Phi\nabla^2\Phi$. Next, using equation (\ref{eq:metric}), the Jordan frame metric
potentials are
\begin{equation}
 \Phi_{\rm J}= \Phi + \alpha\phi -\frac{B^2}{2\Lambda^2}\dot{\phi}^2+\cdots,\quad
\Psi_{\rm J} = \Psi - \alpha\phi -\frac{B^2}{2\Lambda^2}{\phi^\prime}^2+\cdots\label{eq:metpertsJ},
\end{equation}
where the dots denote higher order cross-terms. Now since the Einstein and Jordan frame metric potentials must be of the same
order, it is inconsistent to take $\alpha\phi,\dot{\phi}^2/\Lambda^2,{\phi^\prime}^2/\Lambda^2>\Phi$; we must impose that
these quantities are at least of the same order as the metric perturbations in each frame. In what follows we will denote
quantities that are of the same order or smaller using the notation $\sim$. We hence have the relations\footnote{Note that using
the
identity $\partial_k\phi\partial^k\phi=\partial_k(\phi\partial^k\phi)-\phi\nabla^2\phi$ we have
$\partial_k\phi\partial^k\phi\sim\phi\nabla^2\phi$.}
\begin{align}
 \alpha\phi&\sim\pn,\label{eq:phiords1}\\
\frac{B^2X}{\Lambda^2}&\sim \pn^2,\label{eq:phiords3}\\
X&\sim \pn\nabla^2\pn\quad\textrm{and}\\
8\pi G \rho\mmm\frac{B^2X}{\Lambda^2}&\sim \pn^2\nabla^2\pn. \label{eq:phiord2}
\end{align}

We are now in a position to find the non-relativistic limit of equation (\ref{eq:sfeom}). Ignoring all terms that are
post-Newtonian, we have 
\begin{equation}\label{eq:NR1}
 -\ddot{\phi}+\nabla^2\phi=8\pi \alpha G\rho\mmm+V(\phi)_{,\phi}.
\end{equation}
The non-relativistic limit of this equation corresponds to neglecting time-derivatives\footnote{Consider a system of
mass $M$, and length scale $R$. If the system has a
time-dependence characterised by some frequency $\omega$ then the only dimensionless quantity one can form is
$\Omega^2=\omega^2R^3/GM$, which must be $\mathcal{O}(1)$ if time-derivatives are to be important. One then has $\omega^2\sim
GM/R\cdot 1/R^2\sim \pn\nabla^2\pn$, where $\pn$ is the Newtonian potential. We then have $\ddot{\phi}\sim\pn\nabla^2\phi$
and time-derivatives are hence post-Newtonian compared with spatial ones.} to find
\begin{equation}\label{eq:newtlim}
 \nabla^2\phi=8\pi \alpha G\rho\mmm+V(\phi)_{,\phi}.
\end{equation}
Interestingly, the contributions coming from the disformal part of the metric do not survive in the non-relativistic limit and
equation (\ref{eq:newtlim}) is identical to that found in scalar-tensor theories that contain only a conformal coupling to the
metric.

Finally, note that the Einstein frame density is not conserved, the conserved density is the Jordan frame density $\tilde{\rho}$.
The energy-momentum tensors in the two frames are related by (see \cite{Zumalacarregui:2012us}, appendix A)
\begin{equation}
T^{\nm}\mmm=A^6(\phi)\sqrt{1-\frac{2B^2X}{\Lambda^2}}\tilde{T}\mmm^{\nm}.
\end{equation}
Applying equation (\ref{eq:phiords3}) and recalling that $\rho\mmm\sim\nabla^2\pn$ we can see that the two differ at
post-Newtonian order only and so $\tilde{\rho}\mmm=\rho\mmm$ at
Newtonian order. The Jordan and Einstein frame masses found by integrating over the density are identical at Newtonian order.

\subsubsection{The Fifth-Force}

Since matter moves on geodesics of the Jordan frame metric, a non-relativistic particle's motion is governed by
\begin{equation}\label{eq:JFGD}
 \ddot{x}^i+\tilde{\Gamma}^i_{00}=0,
\end{equation}
where a tilde denotes Jordan frame quantities. Defining
\begin{equation}\label{eq:kdef}
 \mathcal{K}^\alpha_{\mu\nu}=\tilde{\Gamma}_{\mu\nu}^\alpha-\Gamma_{\mu\nu}^\alpha,
\end{equation}
the Einstein-frame equation is\footnote{In fact, this assumes that $\Gamma^0_{00}=0$ so that $\dd^2x^0/\dd\lambda^2=0$, where
$\lambda$ is an affine parameter along the geodesic. Using equation (\ref{eq:kdef}), one finds that $\Gamma_{00}^0$ contains terms
proportional to time derivatives of the field as well as post-Newtonian spatial derivatives. Therefore, the non-relativistic
limit of the geodesic equation in the Einstein frame is given by (\ref{eq:geodde}) only after time-derivatives and post-Newtonian
corrections are ignored.}
\begin{equation}\label{eq:geodde}
\ddot{x}^i +\tilde{\Gamma}^i_{00}=\ddot{x}^i +\Gamma^i_{00}+\mathcal{K}_{00}^i=0.
\end{equation}
Recalling that $\Gamma^i_{00}=\nabla^i\pn$, one can see that the effects of the fifth-force are all contained in
$\mathcal{K}_{00}^i$
or, more specifically, the fifth-force is $F_5^i=-\mathcal{K}^i_{00}$. Using the metric (\ref{eq:metric}), one finds
\begin{equation}\label{eq:f5}
 \vec{F}_5=-\left(1-2\frac{B^2X}{\Lambda^2}\right)^{-1}\left[\alpha+\frac{B^2}{\Lambda^2}\left(\ddot
{\phi}+\left(\gamma-\alpha\right)\dot{\phi}^2\right)\right]\nabla\phi.
\end{equation}
Ignoring all terms that are post-Newtonian one finds
\begin{equation}\label{eq:F5mink}
 \vec{F}_5=-\alpha\nabla\phi,
\end{equation}
which is identical to the force-law found in scalar-tensor theories that include a conformal coupling to matter only. 

\subsubsection{Screening}

The non-relativistic limit of the field's equation of motion is given by (\ref{eq:newtlim}). If one wishes to have the scalar
drive the cosmic acceleration on large scales then one typically has $V_{,\phi}\ll G\rho$ and one is left with the Poisson
equation
\begin{equation}
 \nabla^2\phi=8\pi\alpha G\rho\mmm.
\end{equation}
In this case, one has $F_5=2\alpha^2F_{\rm N}$ and so there are $\mathcal{O}(1)$ fifth-forces on all scales unless one fine-tunes
$\alpha$ to values small enough to evade solar system tests. One exception to this is chameleon-like theories, which use
non-linear but not post-Newtonian effects to reduce the effective source for the field gradient so that the effective scalar
charge of the object is greatly reduced compared with its mass. Another method of screening these theories is to invoke the
Damour-Polyakov effect \cite{Damour:1994zq}, whereby $V(\phi)=0$ and $A(\phi)$ has a minimum so that the cosmological evolution
drives the field towards this so that $\alpha(\phi)\approx0$ in the late time universe. These theories cannot account for dark
energy unless a scalar potential is re-introduced. In this case the screening mechanism will generically be lost unless one
ensures that the minima of both $A$ and $V$ coincide. To date, this scenario has yet to be studied. In either case, no new
screening mechanism is present due to the disformal coupling.

Recently, \cite{Koivisto:2012za,Zumalacarregui:2012us} have presented a new screening mechanism, the \textit{disformal screening
mechanism}, which arises in the theories considered here. There, the condition
$B^2A^2\rho\mmm/\Lambda^2\gg\mpl^2$ is imposed and they find
\begin{equation}\label{eq:disfscreen}
 \frac{B(\phi)^2}{\Lambda^2}\ddot{\phi}=\alpha-\frac{B(\phi)^2}{\Lambda^2}\left(\gamma-\alpha\right)\dot{\phi}^2.
\end{equation}
It is then argued that $\ddot{\phi}\rightarrow0$ is a generic feature of this equation and so time-derivatives are suppressed.
The condition $B^2A^2\rho\mmm/\Lambda^2\gg\mpl^2$ is not sufficient to neglect many of
the terms present in (\ref{eq:sfeom}), the most important being $\nabla^2\phi$. By neglecting this, one has implicitly assumed
that $B^2A^2\rho\mmm\ddot{\phi}\gg\mpl^2\Lambda^2\nabla^2\phi$. For this reason, one cannot self-consistently determine the
validity of the this assumption because one lacks the radial profile. Note also that this equation contains many terms that we
have found to be post-Newtonian and has prioritised them over the leading-order Newtonian ones. The relevant limit of equation
(\ref{eq:sfeom}) is equation (\ref{eq:newtlim}) and we have seen here that this contains no new screening mechanisms due to the
disformal coupling. It is not possible to screen the contribution from the disformal coupling using local dynamics deriving from the coupling 
itself\footnote{By this, we mean that one can add terms to the Lagrangian such as Galileon-like operators or those that exhibit the chameleon effect 
(or other similar mechanisms) but there is no new local screening mechanism that results from the disformal coupling.}. In the next subsection, we 
will see that this conclusion is altered once one accounts for the fact that we
live in a space-time that is asymptotically FRW and not Minkowski.

\subsection{FRW Space-Time}
\label{sec:frwnewtlim}

In this section we will present the equivalent results for the non-relativistic limit of the field equation (\ref{eq:newtlim})
and force-law (\ref{eq:F5mink}) when we take the background to be FRW and not Minkowski. In order to avoid repeating
the calculation above, here we will present only the main results and sketch out the changes that occur. A complete derivation of
these
results is presented in appendix \ref{sec:FRWnewt}. 

One major difference is that FRW space-time is not a vacuum solution of the Einstein equations and so cross-terms appear in the
equation of motion for the inhomogeneous component of the field. In particular, one has
$T^{\nm}\mmm=\textrm{diag}(\rho_\infty(t)+\rho\mmm(r,t),0,0,0)$, where we have included non-relativistic species
only\footnote{One could include relativistic species too, which indeed contribute to the background evolution of the field. Since
we are interested in the non-relativistic limit of inhomogeneous perturbations about this background we have neglected these in
the interest of brevity. Including them will not alter the conclusions of this section.}. The inhomogeneous component of the
density $\rho_\infty$ sources the coupled Friedmann-scalar field system. Writing $\phi=\phi_\infty(t)+\vpp(r,t)$, the homogeneous
component $\phi_\infty$ is sourced by the background equations and the inhomogeneous component $\vpp$ is sourced by the
inhomogeneous density perturbation $\rho\mmm(r,t)$. The reader should note that whereas we refer to inhomogeneous components as
\textit{perturbations}, we are not linearising the equations or treating $\vpp$ as being $\ll\phi_\infty$. Instead, we are
splitting the various quantities into homogeneous and inhomogeneous components and finding the non-relativistic (i.e. weak-field)
forms of the resultant equations (see \cite{Hui:2009kc} for a discussion on this).

We consider the perturbed form of the FRW metric in the conformal Newtonian gauge using the coordinate time so that
\begin{equation}\label{eq:FRWpert}
\dd s^2=\left[-1-2\Psi(r,t)\right]\dd t^2+a(t)^2\left[1-2\pn(r,t)\right]\dd\vec{x}^2.
\end{equation}
The Newtonian limit corresponds to ignoring time-derivatives and terms that are suppressed by factors of the Hubble parameter (for
example, $\nabla^2\phi\gg H^2\phi$ in this limit). Furthermore, since the scale factor evolves over a Hubble-time, we can set
$a(t_0)=1$. Taking the Newtonian limit of equation (\ref{eq:sfeom}) one finds
\begin{equation}\label{eq:newttime}
 \nabla^2\vpp=8\pi G Q\rho\mmm+V_{,\vpp}
\end{equation}
where the local scalar charge is 
\begin{equation}\label{eq:Qdef}
 Q\equiv \frac{\alpha+\frac{B^2}{\Lambda^2}\left(\ddot{\phi}_\infty+\dot{\phi}_\infty^2\left[\gamma-\alpha\right] 
\right)}{1-\frac{B^2\dot{\phi}_\infty^2} {\Lambda^2 } }.
\end{equation} 
The fifth-force is found by taking the non-relativistic limit of the geodesic equation (see appendix \ref{sec:FRWnewt}) to
find
\begin{equation}\label{eq:f5cos}
 \vec{F}_5=-Q\nabla\phi.
\end{equation}
This force can be screened by either suppressing the field gradient or reducing $Q$ to small values such that solar system
constraints are evaded. One could use the chameleon mechanism to achieve this but in this work we are interested in the effects
of the disformal coupling and so we will not investigate this here. Furthermore, one could tune the scalar potential such that
the mass $m_0$ defined by $V_{,\phi}\sim m_0^2\phi$ is large compared with solar system scales giving a short ranged fifth-force.
If we want the scalar field to drive the cosmic acceleration then we expect $m_0\sim H_0$ and so this case is not of interest to
us.

In the case of perturbations about Minkowski space we found that the effective scalar charge is equal to $\alpha$ and so one must
fine-tune this in order to pass solar system tests. In this more realistic case, we have found an additional contribution due to
the cosmological dynamics of the homogeneous component of the field. If the cosmological dynamics are such that 
\begin{equation}
 \alpha(\phi_\infty)\approx-\frac{B^2}{\Lambda^2}\left(\ddot{\phi}_\infty+\dot{\phi}_\infty^2\left[
\gamma(\phi_\infty)-\alpha(\phi_\infty)\right] \right)
\end{equation}
then the local scalar charge is heavily suppressed. We have been unable to find a specific model that realises this
dynamically but this is not to say that one does not exist. This scenario requires a dynamical relation between two \textit{a
priori} unrelated couplings and so one would expect such a relation to be far from generic\footnote{That being said, models where
the
conformal and disformal factor are related arise in string theory scenarios \cite{Koivisto:2013fta} and may be good candidates to
realise such a mechanism.}. This means that when considering generic theories, one generally requires that the conformal factor is not present if one 
wishes
to screen. When this is imposed, one has 
\begin{equation}\label{eq:QnoA}
 Q=\frac{\frac{B^2}{\Lambda^2}\left(\ddot{\phi}_\infty+\gamma(\phi_\infty)\dot{\phi}_\infty^2
\right)}{1-\frac{B^2\dot{\phi}_\infty^2} {\Lambda^2 } }.
\end{equation} 
This is far easier to screen since the local charge is proportional to the first and second time-derivatives of the cosmological
field. Quintessence theories (see \cite{Copeland:2006wr} for a review) drive the acceleration when the field rolls down the
potential and enters a slow-roll phase such that $\ddot{\phi}_\infty\ll3H\dot{\phi}$ and so one
may expect that
$Q\approx0$ is naturally satisfied when $\Lambda$ is large enough without the need for excessive fine-tuning
(beyond that already needed for
quintessence to match the current observations) and without having to contrive specific forms for $V(\phi)$ and $B(\phi)$. We
will see in the next section that this is indeed the case.

It is enlightening to pause to discuss the new features we have derived in this section. A study of
the disformal coupling alone (i.e. setting $A(\phi)=1$ from the outset) would n\"aively give $\nabla^2\phi=0$ in
the non-relativistic limit since one generally assumes Minkowski space and so one would be led to argue that local tests of
gravity do
not apply because scalar field gradients are not sourced. Indeed, this is what is argued in \cite{Noller:2012sv}. What we have
shown here is that the cosmological dynamics are
important because they source spatial gradients on small scales. This effect is also seen in galileon theories of gravity, where
the expansion about Minkowski space leads to a purely static situation but the covariantisation leads to time-variations in $G$
in the solar system \cite{Barreira:2013xea}.

\subsection{The Decoupling Limit}\label{sec:dec}

Finally, before proceeding to study the cosmology of these theories, we pause to discuss the relation between disformally coupled
metrics and the \textit{disformal coupling} that arises in the decoupling limit of massive gravity
\cite{deRham:2010kj,Hinterbichler:2011tt}\footnote{Note that since we are including generalised couplings $A(\phi)$ and $B(\phi)$
the Lagrangian presented here is not quite the same as the one arising in massive gravity.} and galileon theories
\cite{Nicolis:2008in}:
\begin{equation}\label{eq:disflag}
 \frac{\mathcal{L}}{\sqrt{-g}}\supset -\frac{1}{2}\nabla_\mu\phi\nabla^\mu\phi+\cdots+\cdots T\ln
A(\phi)+\frac{B^2(\phi)}{2\Lambda^2}T^{\mu\nu}\nabla_\mu\phi\nabla_\nu\phi,
\end{equation}
where $\cdots$ include other matter species as well as other self-interactions of the field. Indeed, this is the Lagrangian (with
$B(\phi)$=1 in our conventions) studied by \cite{Kaloper:2003yf,Brax:2014vva,Brax:2014zba} who
calculate the force between two particles to one-loop. Now the first term in (\ref{eq:disflag}) gives the correct contribution to
the field's equation of motion around any background but one cannot reproduce the disformal contributions to (\ref{eq:sfeom})
using the second and so one should really think of this as an effective Lagrangian describing the theory on a fixed background.
One important question to answer is: Is this the Minkowski space effective theory for disformally coupled metrics? Indeed, one
can also obtain (\ref{eq:disflag}) using a covariantisation of the form $B(\phi)G_{\mu\nu}\nabla^\mu\phi\nabla^\nu\phi$
\cite{Appleby:2011aa} but it is not necessarily the case that the covariantisation is unique. Setting $A=1$, the equation of
motion is
\begin{equation}\label{eq:intdisf}
 \mpl^2\nabla^2\phi+\cdots-\nabla_\mu\left(\frac{B(\phi)}{\Lambda^2}
T^\nm\nabla_\nu\phi\right)+\frac{\gamma(\phi)B^2(\phi)}{\Lambda^2}T^\nm\nabla_\mu\phi\nabla_\nu\phi=0 ,
\end{equation}
where $\cdots$ indicate terms coming from non-disformal parts of the Lagrangian. A non-relativistic source has
$T^\nm=\rho\mmm u^\nu u^\nu$ with $u^\mu=(1,0,0,0)$ and $\dot{\phi}=0$ and so there is no disformal contribution to the equations
of motion. Indeed, this is what we found in equation (\ref{eq:newtlim}). This implies that there is no fifth-force in the static,
Minkowski limit. Indeed, \cite{Brax:2013nsa} argue that this must be the case. A coupling of the form
$T\mmm^\nm\nabla_\mu\phi\nabla_\nu\phi$ involves two scalars coupled to two matter particles and so the contribution from the
scalars to
the 2-2 scattering amplitude must appear at loop level and hence there are no classical contributions to the force from this
interaction.

Next, consider the case where $\phi$ has some background VEV $\phi_\infty$ such that $\phi=\phi_\infty+\vpp$. In this case one
finds an interaction of the form $\mathcal{L}/\sqrt{-g}\supset2
B(\phi_\infty)T^\nm\nabla_\mu\phi_\infty\nabla_\nu\phi/\Lambda^2+\cdots$.
Since $\phi_\infty$ is a classical background VEV, there is now a three-point interaction involving one scalar and two matter
particles and so it is possible to form a tree-level diagram for the 2-2 scattering of matter particles mediated by a single
scalar. Indeed, performing this split in equation (\ref{eq:intdisf}) and taking $\phi_\infty=\phi_\infty(t)$ one
finds\footnote{Note that $\dot{\rho}\mmm=0$ since we have argued above that any non-conservation of the energy-momentum tensor is
a post-Newtonian effect.}
\begin{equation}
\nabla^2\phi+\cdots=8\pi GQ\rho\mmm,
\end{equation}
where $\vpp=\vpp(\vec{x})$ as is appropriate for the non-relativistic limit. This is precisely equation (\ref{eq:newttime}) and
so we see that (\ref{eq:intdisf}) is indeed the effective Lagrangian for disformally coupled metric expanded about flat space
with both a zero and time-dependent VEV. 

\section{Cosmology}\label{sec:cosmo}

In this section we investigate the cosmology of disformally coupled theories with a focus towards examining the screening of the
fifth-force using the cosmological dynamics discussed in the previous section. The Friedmann equations are identical to those
found in GR:
\begin{align}
 3H^2&= 8\pi G\rho+\frac{\dot{\phi}_\infty^2}{2}+V(\phii)\label{eq:fried1}\\
\dot{H}&=-4\pi G\rho -\frac{\dot{\phi}_\infty^2}{2}.\label{eq:fried2}
\end{align}
This is because we are working in the Einstein frame. All of the modifications appear in the equations of
motion \cite{Zumalacarregui:2012us}:
\begin{align}
 \ddot{\phi}_\infty+3H\dot{\phi}_\infty+V(\phii)_{,\phi}&=-Q_0\quad\textrm{and}\label{eq:fired3}\\
\dot{\rho}\mmm+3H\rho\mmm=Q_0\dot{\phi}_\infty
\end{align}
 where
 \begin{equation}\label{eq:Q0}
  Q_0=8\pi G\rho\frac{\alpha+\frac{B^2}{\Lambda^2}\left(\left[\gamma-\alpha\right]
 \dot{\phi}_\infty^2-3H\dot{\phi}_\infty-V_{,\phi}\right)}{1+\frac{B^2}{\Lambda^2}\left(8\pi G
 \rho-\dot{\phi}_\infty^2\right)}.
\end{equation}
One can then define the density parameter and effective equation of state for the field:
\begin{align}
\Omega_\phi&= \frac{\dot{\phi}_\infty^2}{6H^2}+ \frac{V(\phi)}{3H^2}\label{eq:omphi}\quad\textrm{and}\\
\omega_\phi &= \frac{\dot{\phi}^2_\infty-2V}{\dot{\phi}^2_\infty+2V}\label{eq:wphi}.
\end{align}
These are identical to the expressions found in quintessence models and so these theories drive the acceleration of the universe
in an identical manner to quintessence, the field enters a slow-roll-like phase and $\omega_\phi\approx1$. What can differ is
the field dynamics.

Now we have argued in the previous section that models with both a conformal and disformal factor will not generally screen and so
from here on we will set $A(\phi)=1$. We also argued that any model where the cosmological time-derivatives are small at late
times should be screened locally and so in order to demonstrate this, we will work with the simplest model that we expect to
realise this:
\begin{equation}\label{eq:cocnmod}
 V(\phi)=m_0^2(1+\frac{\lambda_0^2}{2}\phi^2)\quad\textrm{and}\quad B(\phi)=1.
\end{equation}
This form of the disformal factor gives the local scalar charge:
\begin{equation}\label{eq:Qspec}
 Q=\frac{\frac{\ddot{\phi}_\infty}{\Lambda^2}}{1-\frac{B^2\dot{\phi}_\infty^2}{\Lambda^2 }}.
\end{equation}
One would expect the field to be
over-damped in the early universe until $H\sim \lambda_0m_0$, at which time the field will roll down the potential towards the
minimum. Near the minimum, the field will slow-roll and one expects $\dot{\phi}_\infty^2\ll V(\phii)$. More importantly, if
$\ddot{\phi}_\infty\ll\Lambda^2$ then $Q\ll 1$ and all fifth-forces will be screened in the solar system. We will take
$m_0\sim H_0$ in order to produce the correct vacuum energy density today\footnote{Assuming no other contributions from other
sectors of the universe. We will not have anything to say about the old cosmological constant problem here.}. If $\lambda_0m_0\gg
1$ then we expect this to happen far in the past so that the system looks like a cosmological constant today. In the opposite case where $\lambda_0\ll 
1$
we
expect the field to be over-damped today so we again have $Q=0$ locally. The intermediate case is $\lambda_0\approx1$. In this
case the field has begun to roll sometime in the recent past and we expect to have some small but non-zero local scalar charge.
For this reason, we will focus on models where $\lambda_0\sim1$\footnote{That is not to say that other models are not
interesting but they cannot be probed using solar system or laboratory experiments since they predict zero scalar charge
locally.}.

\begin{figure}[ht]\centering
\includegraphics[width=0.5\textwidth]{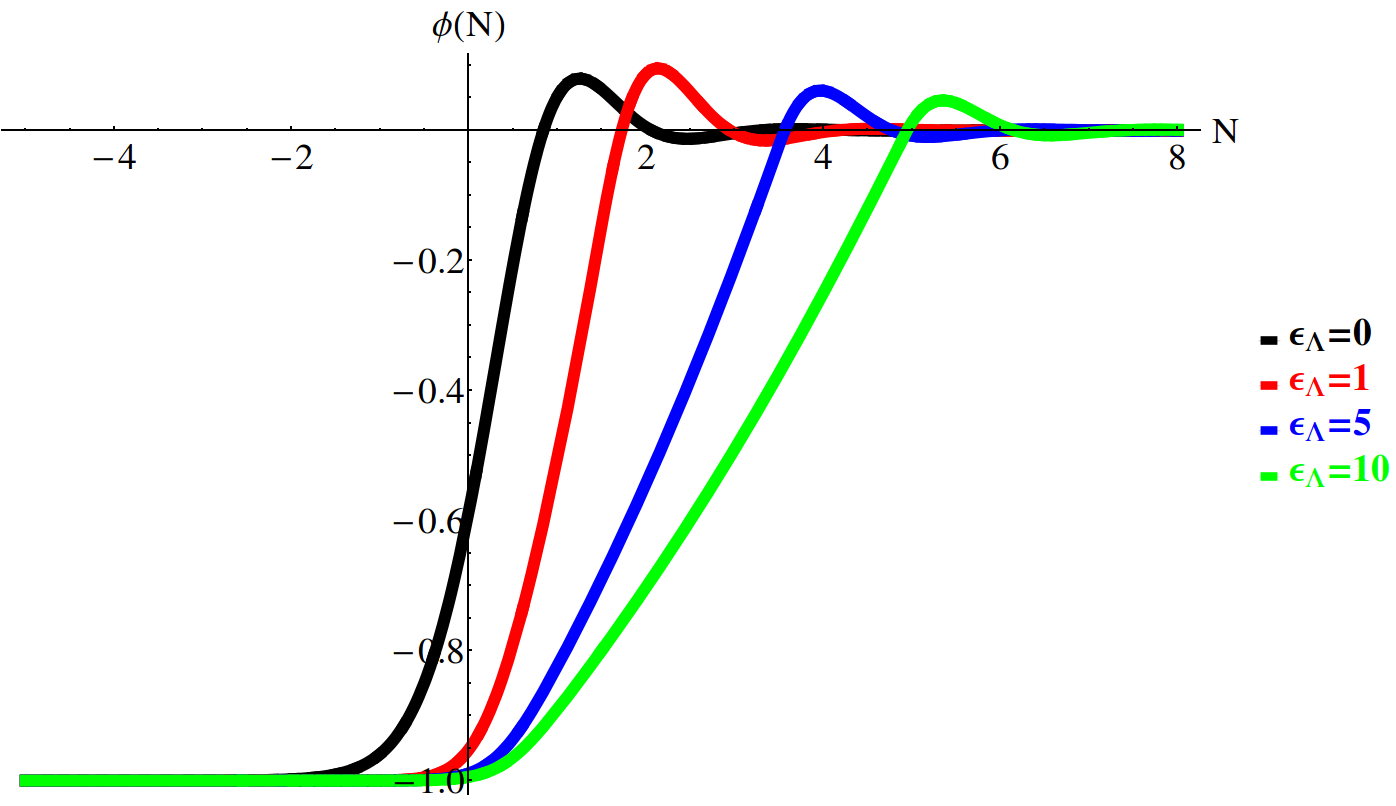}
\caption{$\phii(N)$ as a function of $N$ for the models indicated in the legend.}\label{fig:phiN}
\end{figure}

In order to verify this, we will integrate the Friedmann-scalar field system numerically; We will work not with the coordinate
time t but with $N \equiv \ln a(t)$ and use a prime to denote derivatives with respect to $N$. We begin by introducing the
dimensionless quantities
\begin{equation}\label{eq:cosqaunts}
 \Omega\mmm\equiv \frac{8\pi G\rho\mmm}{3H^2},\quad\epsilon_m\equiv \left(\frac{m_0}{H_0}\right)^2,\quad
\epsilon_{\Lambda}\equiv\left(\frac{H_0}{\Lambda}\right)^2\quad\textrm{and}\quad\epsilon_H=\left(\frac{H_0}{H}\right)^2.
 \end{equation}
In terms of these quantities we have
\begin{equation}\label{eq:Q0quan}
 Q_0=-3\Omega\mmm\frac{\lambda_0\epsilon_\Lambda\epsilon_m\phii+3\frac{\epsilon_\Lambda}{\epsilon_H}{\phii^\prime}^2}{1+\frac{
\epsilon_\Lambda } { \epsilon_H } \left(3\Omega\mmm- { \phii^\prime } ^2\right)}
\end{equation}
and the Hubble constraint is
\begin{equation}
 \epsilon_H=\frac{6-6\Omega\mmm-{\phii^\prime}^2}{\lambda_0\epsilon_m\phii^2}.
\end{equation}
Fixeing $\lambda$ and $\epsilon_m$ such that the observed dark energy density today agrees roughly with the observed value 
$\Omega_{\rm DE}\approx0.7$, the only free parameter 
is $\epsilon_\Lambda$, which controls the size of the disformal effects
on the dynamics. We will take $\phii(N_i)=-1$ and $\phii^\prime(N_i)=0$ as our initial conditions and begin the evolution at
$N_i=-5$. We do not include radiation since our aim here is not to produce a realistic cosmology but rather to study the effects
of the disformal coupling on the cosmological dynamics and local scalar charge at late times. In figure
\ref{fig:phiN} we plot the evolution of the field as a function of time. The case with $\epsilon_\Lambda=0$ is the equivalent
quintessence model. One can see from the figure that stronger disformal couplings delay the point at which the field begins to
roll to increasingly later times. 

\begin{figure}[ht]\centering
\subfigure{\includegraphics[width=0.45\textwidth]{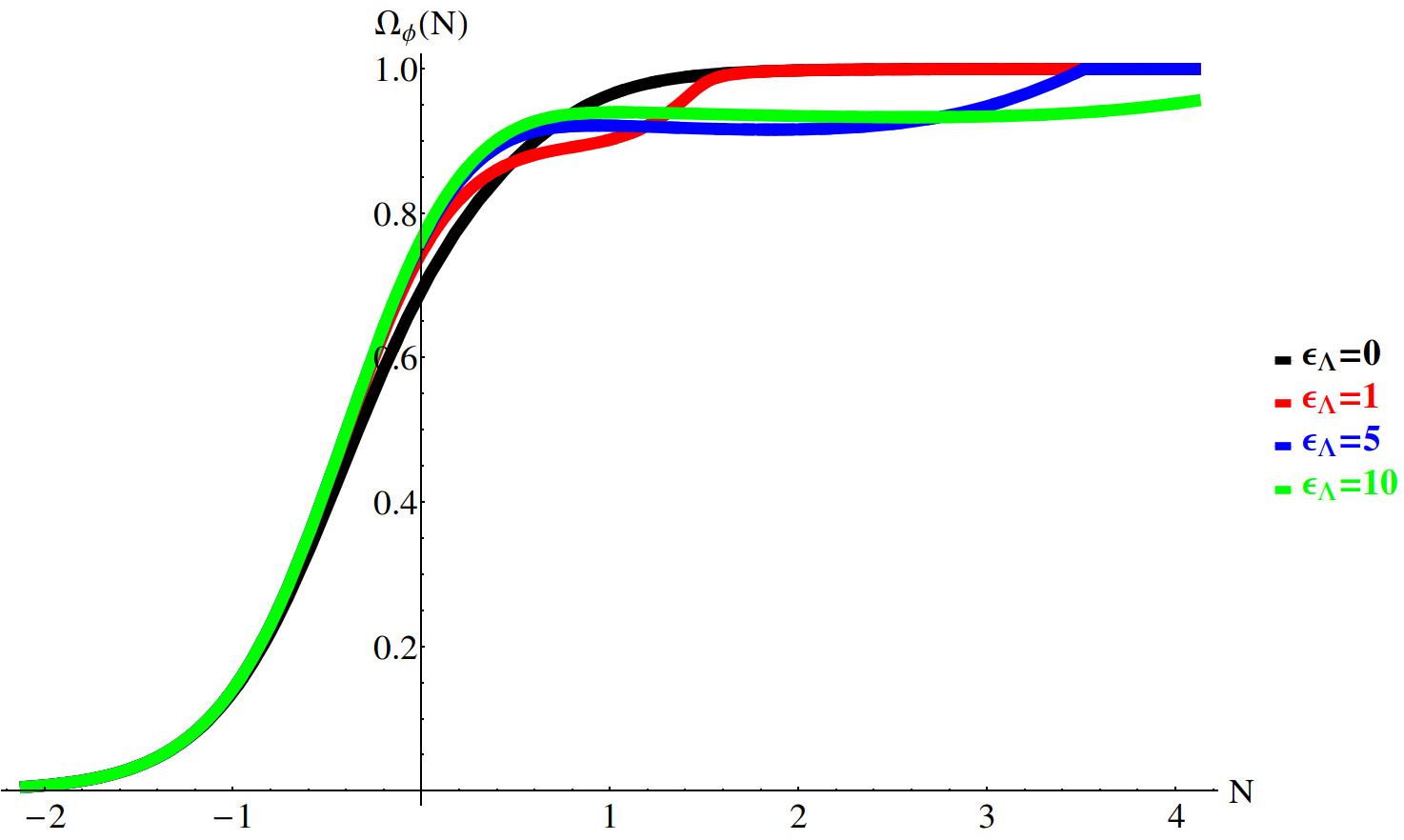}}
\subfigure{\includegraphics[width=0.45\textwidth]{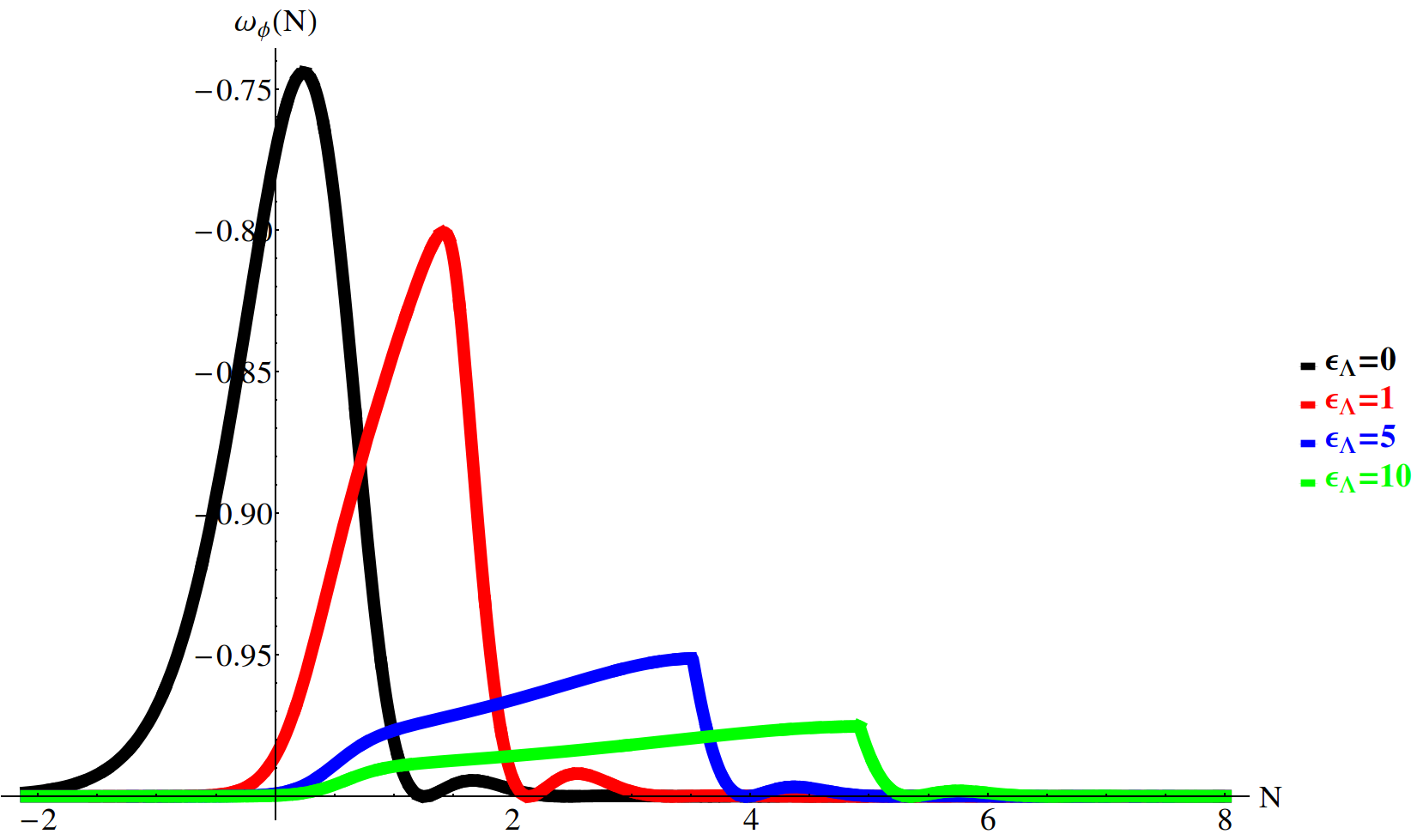}}
\caption{$\Omega_\phi$ and $\omega_\phi$ as a function of $N$ for the models indicated in the legend.}\label{fig:omegas}
\end{figure}

Next, we plot the evolution of $\Omega_\phi$ and $\omega_\phi$ in figure \ref{fig:omegas}.
One can see the effect of the disformal coupling is to push the onset of dark energy domination to later times. Furthermore,
$\omega_\phi$ behaves qualitatively as expected, it is $-1$ in the far past when the field is over-damped and approaches $-1$
when the field hits its minimum but there is a small bump to larger values when the field is rolling.

\begin{figure}[ht]\centering
\includegraphics[width=0.5\textwidth]{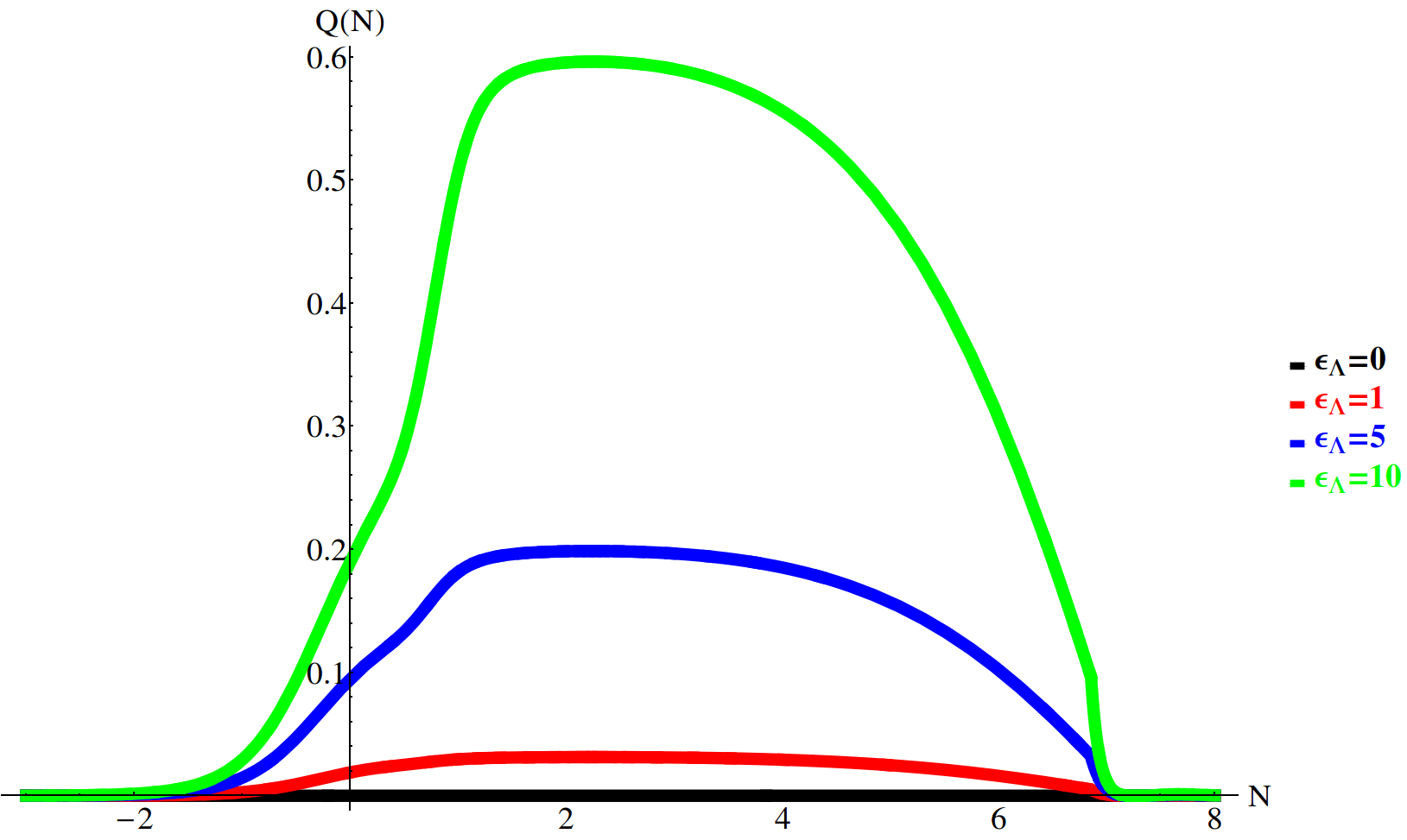}
\caption{$Q(N)$ as a function of $N$ for the models indicated in the legend. The black line has no disformal
coupling and hence $Q$ is identically zero.}\label{fig:QN}
\end{figure}

Finally, we plot the local scalar charge in figure \ref{fig:QN}.
One can see that stronger disformal couplings give larger values of $Q$ and hence larger fifth-forces on small scales. Note also
that the peak of the charge is in the far future, and so even if one finds that fifth-forces are negligible today, models similar
to this one predict that this may not necessarily be the case at later times. Finally, note that the charge is close to zero in
the past
when the field was not rolling. \cite{Zumalacarregui:2012us} studied a model with an exponential potential, which has the field
rolling at all times. They found that the model's predictions for linear cosmological probes such as the CMB and the cold dark
matter power spectrum differed greatly from the $\Lambda$CDM prediction. This may be due to large time-derivatives in the past,
which models with a minimum do not suffer from.

\section{Local Tests}
\label{sec:tests}
We have seen in the previous section that we expect a non-zero scalar charge in the solar system if the mass of the field is of
order Hubble. In this section, we use local tests of gravity to place constraints on the model parameters. Local tests of
gravity generally fall into two classes. Solar system tests probe the structure of gravity using the properties of
spherical objects such as stars and planets and laboratory tests tend to measure the force between objects in either spherical
or planar configurations. For this reason, it will be useful to derive the field profile for both a spherical body and two
plane parallel plates. We will ultimately find that laboratory tests are not good probes of these models. For this reason,
we present
the derivation for the parallel plate configuration in appendix \ref{sec:plates} and quote only the salient results here.

\subsection{Field Profile for a Spherical Object}

We begin by solving equation (\ref{eq:newttime}) for a spherical object of radius $R$. Adopting the model defined in equation
(\ref{eq:cocnmod}), we have
\begin{equation}\label{eq:specificsphere}
 \nabla^2\vpp=m_0^2\vpp+8\pi G Q\rmm,
\end{equation}
where $Q$ is given by equation (\ref{eq:Qspec}) and we evaluate it at the present time so that
$Q=\epsilon_\Lambda\phii^{\prime\prime}$. Since we want
the field to drive the cosmic acceleration, we have $m_0R\ll1$ inside any astrophysical object of interest and so the solution of
equation (\ref{eq:specificsphere}) is
\begin{equation}\label{eq:sphereinside}
 \vpp(r)=\vpp_c+2Q\pn(r)\quad r<R,
\end{equation}
where $\pn$ is the Newtonian potential sourced by the object and $\phi_c$ is the field value at the centre. Since the
strength of the fifth-force depends on $\vpp^\prime$ only we will not calculate its value. Outside, we have $\rmm=0$ and the
solution is
\begin{equation}\label{eq:sphereoutside}
 \vpp(r)=-\frac{C}{r}e^{-\mu r}\quad r>R,
\end{equation}
where $\mu^2\equiv\lambda_0 m_0^2$, $C$ is an integration constant and we have set $\vpp(r\rightarrow\infty)=0$. Using the fact
that $\dd\pn/\dd r=GM/r^2$ at $r=R$ and the requirement that the
field gradient be continuous across the boundary we have
\begin{equation}
2\frac{GQM}{R^2}=-\frac{C}{R^2}e^{-\mu R}\left[1+\mu R\right].
\end{equation}
Since $\mu R\ll1$ we have $C\approx-2Q GM\exp(\mu R)$ so that
\begin{equation}\label{eq:sphere}
 \vpp(r)\approx2Q\frac{GM}{r}e^{-\mu (r-R)}\quad r>R.
\end{equation}
In chameleon-like theories, one has $Q\ll 1$ due to the non-linear screening. In disformal theories, this is achieved whenever
the cosmological field is slowly rolling. Note that this solution is perfectly consistent with the assertions we made when
constructing the non-relativistic limit of the field equations in section \ref{sec:nrlim}, i.e. $\phi\sim\pn$. In fact, it is a
self-consistency check.

\subsection{Solar System Tests}

In this section we will consider two solar system tests of gravity. Light bending by the Sun as measured by the Cassini probe and
lunar laser ranging (LLR).

\subsubsection{The Cassini Probe}
\label{sec:cass}
The Cassini probe measures the PPN parameter $\gamma_{\rm PPN}$---defined as $\gamma_{\rm PPN}=\Psi_{\rm J}/\Phi_{\rm J}$ in the
Jordan frame---by measuring light bending by the Sun. Using equation (\ref{eq:metric}), we have
\begin{align}
 \Psi_{\rm J}&=\pn\left(1-\frac{B^2(\phi){\phi^\prime}^2}{2\pn\Lambda^2}\right)\quad\label{eq:self2} \textrm{and}\\
\Phi_{\rm J}&=\pn
\end{align}
so that 
\begin{equation}\label{eq:gamgen}
 |\gamma_{\rm PPN}-1|=\frac{B(\phi){\phi^\prime}^2}{2\Lambda^2\pn}.
\end{equation}
Using the spherical profile obtained in the previous section, we have\footnote{Note that we are taking the impact parameter to be
the solar radius. The minimum impact parameter is $1.6R_\odot$ and so this is a valid approximation.}
\begin{equation}\label{eq:gamspec}
 \frac{B(\phi){\phi^\prime}^2}{2\Lambda^2}= \frac{2Q^2G^2M^2_\odot}{\Lambda^2R_\odot^4}=\frac{2Q^2{\pn^{\odot}}^2}{\Lambda^2
R_\odot^2}.
\end{equation}
Note that (\ref{eq:self2}) implies that $B^2{\phii^\prime}^2/\Lambda^2\sim\pn^2$, which is consistent with the assertion we made
when finding the
non-relativistic limit in section \ref{sec:nrlim} and is therefore another self-consistency check. The Cassini satellite has set
an upper bound on $\gamma_{\rm PPN}$:
$|\gamma_{\rm PPN}-1|\lsim 2\times10^{-5}$
\cite{Bertotti:2003rm}, which imposes the constraint
\begin{equation}\label{eq:cassboundspec}
 \frac{Q^2}{\Lambda^2R_\odot^2}\lsim 5.
\end{equation}
One can see that the Cassini measurement constrains a combination
of $\Lambda^2$ and the
cosmological time-derivatives of the field. For this reason, a careful analysis of the constraints imposed by this bound would
require one to use observations to find the best-fit cosmological model and use this to constrain the parameter $\Lambda$. This is
beyond the scope of this work but we can find order-of-magnitude constraints if we assume that the cosmological evolution in the
recent past does not differ drastically from $\Lambda$CDM. This is a reasonable assumption since the expansion is driven by a
quintessence-like mechanism and we have seen in section \ref{sec:cosmo} that the evolution of the system is very similar to the
quintessence case but delayed to later times. Translating the bound into our model parameters, one finds that the resultant
expression---we do not present it here due to its cumbersome form---can be cast in terms of the dimensionless
ratios $V/H_0^2$, $V_{,\,\phi}/V$ and $\dot{\phi}\inff/H_0$. For this reason, one does not need to specify $\epsilon_m$ or
$\lambda_0$ in order to place constraints but three individual measurements are required. For this reason, we use the PLANCK dark
energy measurements of the of the
$\omega_0$--$\omega_a$ parametrisation $\omega(a)=\omega_0+\omega_a(1-a)$ including baryon acoustic oscillations\footnote{Since
we are not interested in precision tests but rather order-of-magnitude constraints, the choice of data set is largely irrelevant
and we have made this choice in order to provide concrete numbers for the analysis.} $\omega_\Lambda=0.699$, $\omega_0=-0.98$ and
$\omega_a=-0.39$. Using these, one can use equations (\ref{eq:fried1})--(\ref{eq:fired3}) to calculate
${\phi^{\prime\prime}}$. Finally, using $H_0R_\odot/c\approx5\times10^{-18}$ we find
\begin{equation}\label{eq:casscons}
 \frac{\Lambda}{H_0}\gsim 4.24\times10^5.
\end{equation}
This gives the value of $Q$ locally as $|Q|<4.9\times10^{-12}$. Given that we expect $\mu R\ll1$ where $R$ is any typical
astrophysical length scale, the theory behaves like the Newtonian limit of general relativity with an effective value of Newton's
constant with $\Delta G/G\equiv 2Q^2\sim 10^{-23}$. 

\begin{figure}[ht]\centering
\includegraphics[width=0.5\textwidth]{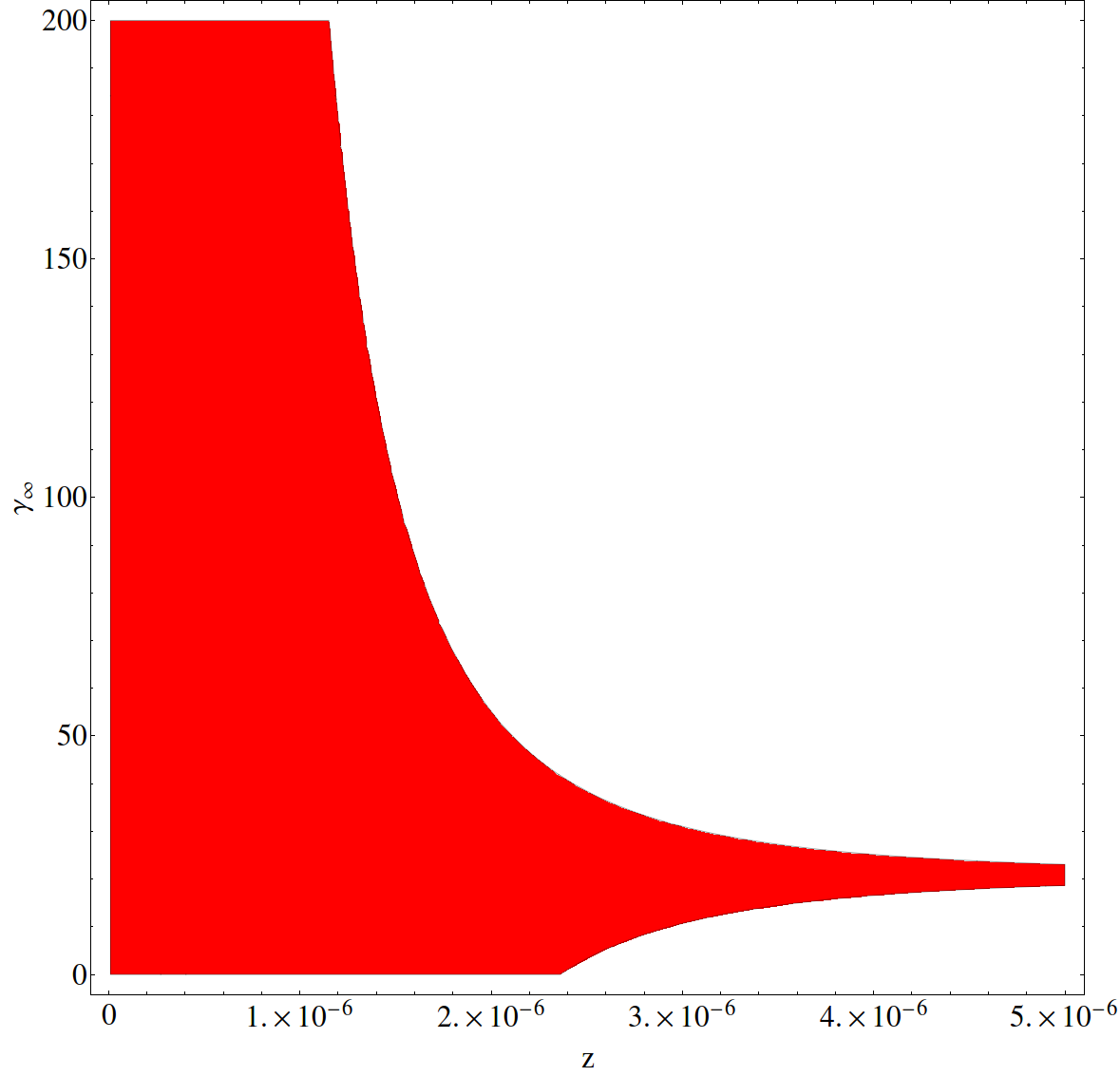}
\caption{The region in the $z$--$\gamma_\infty$ plane (red) where the Cassini constraint is
satisfied.}\label{fig:cg}
\end{figure}

One may wonder whether more general models can circumvent this bound. In this case, $Q$ is given by equation
(\ref{eq:QnoA}) with $\gamma=\gamma_\infty\equiv\gamma(\phi_\infty)$ and $B(\phi)=B_\infty\equiv B(\phii)$. Furthermore, the profile for a spherical 
object is the same\footnote{This assumptions ignores any potential non-linear effects
that may cause
$B(\phi_\infty+\vpp)$ to differ from $B_\infty$ greatly. In particular, models that do not have well-defined Taylor expansions
such as $B(\phi)=\phi^{-n}$ require more care. We will not consider these here.} but the bound
(\ref{eq:cassboundspec}) is
\begin{equation}\label{eq:cassboundgen}
 \frac{B_\infty^2Q^2}{\Lambda^2R_\odot^2}\lsim 5.
\end{equation}
We can then use this bound to constrain regions in the $z$--$\gamma_\infty$ plane, where $z=B_\infty H/\Lambda$. This is shown in
figure \ref{fig:cg}. One can see that for $\gamma_\infty\sim 50$ the constraints are tighter
but in general the bound (\ref{eq:casscons}) holds, at least to the same order-of-magnitude when
$B_\infty\sim\mathcal{O}(1)$. When $B_\infty$ differs greatly from this, one can strengthen(weaken) the bound on
$\Lambda$ if $B_\infty\gg1$($B_\infty\ll1$). A full analysis of this requires one to find the best-fit cosmology for a variety of
different models, which is beyond
the scope of this work but we note that $\Delta G/G\equiv 2Q^2\sim 10^{-23}$ holds even when $\gamma$ is allowed to vary unless
one pushes it to large values ($\gsim\mathcal{O}(10^5)$).

\subsubsection{Lunar Laser Ranging}

LLR tests gravity using three separate effects: The Nordvedt Effect, which measures the difference between the free-fall
acceleration of the Earth and Moon towards the Sun due to possible equivalence principle violations; Deviations from the
inverse-square law at distances comparable to the Earth-Moon separation ($a\sim 10^8$ m); and the time-variation of $G$. Since $Q$
does not depend on the structure or composition of the object there are no equivalence principle violations and the first test
is satisfied. Similarly, we have $\mu\sim H_0$ so the range of the fifth-force is larger than the distance between the Earth and
the Moon and so the force is essentially inverse-square at these distances and this test will not yield strong
constraints. Now since the scalar charge is proportional to time-derivatives
of the cosmological field value, one has $\dot{Q}\ne0$ so the third test can be used to place constraints.

\begin{figure}[ht]\centering
\includegraphics[width=0.5\textwidth]{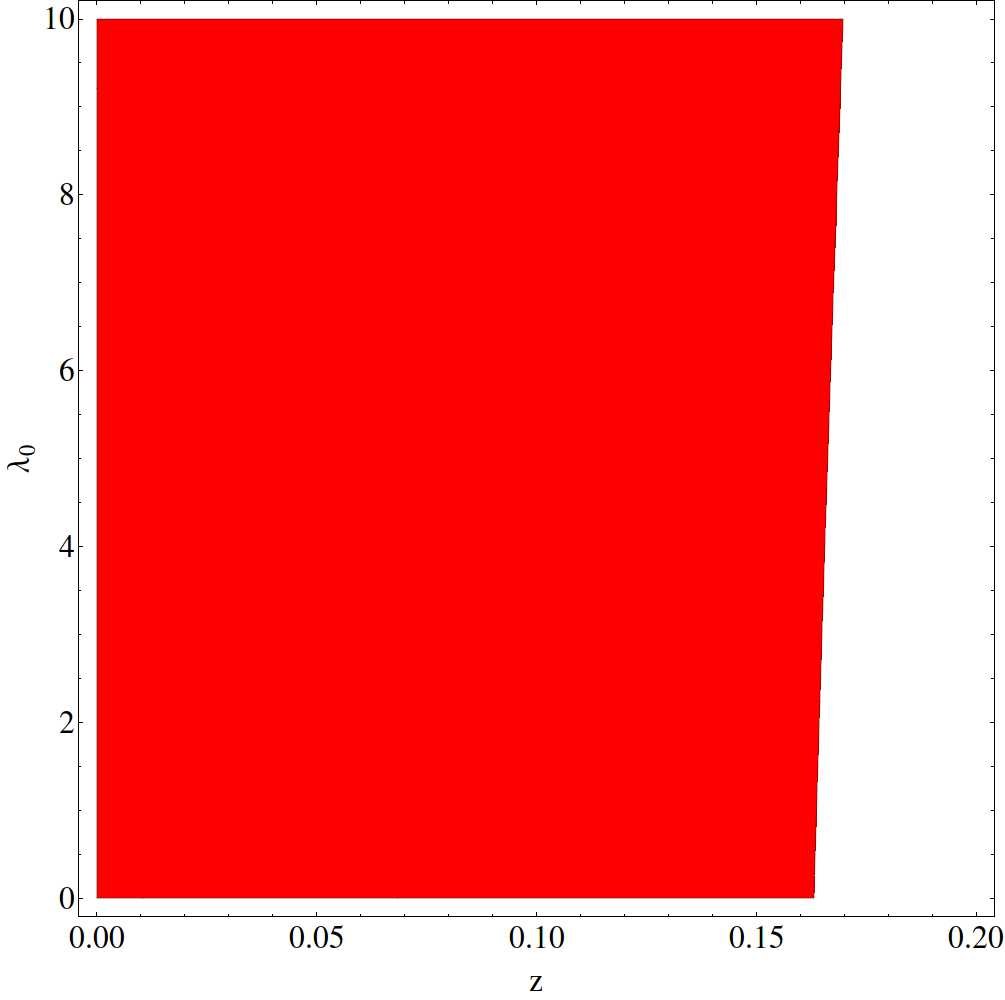}
\caption{The region in the $z$--$\lambda_0$ plane (red) where the LLR constraint from the time-variation of
$G$ is satisfied.}\label{fig:gd}
\end{figure}

Since $a\mu\ll1$, we can ignore the exponential in (\ref{eq:sphere}), in which case one has $G_{\rm
eff}(t)=G(1+2Q^2)$ so that
\begin{equation}
 \frac{\dot{G}\eff}{G}=\frac{4Q\dot{Q}}{1+2Q^2}\approx 4Q\dot{Q}
\end{equation}
where we have assumed that $2Q^2\ll1$, which corresponds to the screened scenario. \cite{Turyshev:2006gm} report
that $\dot{G}/G<6\times 10^{-13}$ yr$^{-1}$, which imposes the constraint $|Q\dot{Q}|<1.5\times10^{-13}$ yr$^{-1}$. For our
specific
model, this gives
\begin{equation}\label{eq:LLRgdotcons}
 \left\vert\frac{\dddot{\phi}_\infty\ddot{\phi}_\infty}{\Lambda^4}\right\vert<1.5\times10^{-13}\textrm{ yr}^{-1},
\end{equation}
which can be written as
\begin{equation}
 \frac{QQ^\prime}{1+2Q^2}<\frac{1.5\times10^{-13}\textrm{ yr}^{-1}}{H_0}\approx 0.002.
\end{equation}
We can translate this into a constraint by taking one time-derivative of (\ref{eq:fired3}) and using a procedure similar to that
described in section \ref{sec:cass}. As was the case with the Cassini bound, the resultant expression is cumbersome and we do not
present it here. The important difference is that the third time-derivative of $\phii$ in equation (\ref{eq:LLRgdotcons}) depends
on four dimensionless ratios---$V/H_0^2$,
$V_{,\,\phi}/V$, $\dot{\phi}\inff/H_0$ and $V_{,\,\phi\phi}/V$---and not three as
was the case with the Cassini bound. The fourth ratio is fixed
uniquely by specifying $\lambda_0\epsilon_m$. In what follows, we will fix $\epsilon_m=1$ so that the field begins to roll around the present epoch 
and treat $\lambda_0$ as the only free parameter. We can then plot the excluded region in the $z$--$\lambda_0$ 
plane, where $z=H/\Lambda$. In practice, we expect $\lambda_0\sim\mathcal{O}(1)$ i.e. $\mu \sim
H_0$. If it is much larger then the
cosmological field will have reached the minimum of the potential today and we would have $Q=0$ locally; If it is smaller, the
cosmological field will be over-damped and again $Q=0$ (see figure \ref{fig:QN}). Therefore, we focus on the region
$\lambda_0\sim\mathcal{O}(1)$. The
region in the $z$--$\lambda_0$ plane where the constraint is satisfied is shown in figure \ref{fig:gd}. One can see
that this test imposes the bound 
\begin{equation}\label{eq:gdotbound}
 \frac{\Lambda}{H_0}\gsim6,
\end{equation}
which is weaker than the one coming from the Cassini probe (\ref{eq:casscons}) by five orders-of-magnitude. Unlike the case of
the Cassini probe, we cannot extend the analysis here to more general models since the time-derivative of $Q$ requires one to
find $\dot{\gamma}$, which is highly model-dependent, although given a specific model the bound can always be calculated.

\subsection{Laboratory Tests}

Chameleon \cite{Mota:2006fz,Brax:2008hh,Brax:2007vm,Brax:2009bk} and galileon
theories \cite{Brax:2011sv} have been constrained using the E\"{o}t-Wash experiment and Casimir force searches. Here, we will
describe each of these in turn.

The E\"{o}t-Wash experiment \cite{Kapner:2006si} is a torsion-balance experiment designed to test deviations from the
inverse-square law. It consists of two circular plates suspended above one another in a configuration such that there is zero
torque on the system if the force between them is exactly inverse-square. The set-up probes distances $d$ of order $55$ $\mu$m to
$9.53$ mm. Since we are considering disformal fields that can drive the cosmic acceleration we have $\mu d\ll1$ for this range
and so deviations from the inverse-square law are negligible and this experiment is not sensitive to the modification of gravity.
Even older experiments that have measured deviations in the range $2$--$105$ cm are satisfied \cite{Hoskins:1985tn}. One may then
wonder if we can constrain other models if we drop the restriction that the field's mass be of order $H_0$. Indeed, chameleon
theories can be probed using this
experiment because the mechanism operates by increasing the field's mass in the solar system by several orders-of-magnitude.
In theory, theories with $\mu\gg H_0$ could indeed be probed using this experiment, however, the cosmological field
would have reached its minimum long before the present epoch so that all cosmological time-derivatives are zero and hence there is
no local scalar charge today. For this reason, one would expect no fifth-force to be sourced in the solar system and hence these
theories cannot be probed using local tests.

Casimir force tests are more promising. They are not designed to test deviations from the inverse-square law but rather measure
the pressure directly and can hence probe the local charge $Q$. Unfortunately, these tests are still not suited to testing
disformal theories of gravity. Casimir force experiments operate in the range $d<10$ $\mu$m where the Casimir
pressure dominates over the Newtonian one \cite{Lambrecht:2011qm}. Due to the thin shell effect, chameleon forces scale with
varying powers of the inverse separation in a similar manner to the Casimir force but, as discussed in appendix \ref{sec:plates},
the disformal force between two plates is simply the Newtonian one with $\Delta G=2Q^2G\ll G$. For this reason, any deviations
from the predicted Casimir force due to disformal couplings are negligible and therefore these experiments cannot probe these
theories\footnote{Note that \cite{Brax:2014vva} were able to obtain a constraint from Casimir force experiments. This is
because the one-loop force is proportional to $r^{-7}$, which is best probed on short length scales such as those used in
these experiments.}.

Finally, note that the conclusions of this section are predicated on the assumption that the field only begins to roll when
$H\sim m_0$ and hence the mass today must be of order Hubble. This is correct for quintessence but we have seen in figure
\ref{fig:phiN} that the effect of the disformal coupling is to delay the time at which the field begins to roll. It is then
possible to construct models where the force is short-ranged but the cosmological field does not begin to roll until the current
epoch due to the effects of the disformal coupling. This is an interesting possibility and certainly merits future attention. 

\section{Comparison with other Constraints}

Recently, \cite{Kaloper:2003yf,Brax:2014vva,Brax:2014zba} have derived the one-loop contributions to various laboratory, astrophysical and
stellar processes from the effective Lagrangian
\begin{equation}
 \frac{\mathcal{L}}{\sqrt{-g}}=\frac{T\mmm^{\mu\nu}\partial_\mu\phi\partial_\nu\phi}{\mathcal{M}^4},
\end{equation}
which, as we showed in section \ref{sec:dec}, gives our equations of motion around Minkowski or FRW space to Newtonian order and
hence we can identify
\begin{equation}\label{eq:M}
 \mathcal{M}^4=2\frac{\mpl^2\Lambda^2}{B_\infty^2}.
\end{equation}
The strongest constraints on $\mathcal{M}$ come from collider searches for beyond the standard model particles and impose
$\mathcal{M}\gsim
10^2$ GeV\footnote{It should be noted that \cite{Brax:2014vva} assume that the scalars are massless whereas ours have a mass of
order Hubble. Since $H_0/E\ll1$ ($E$ is the typical centre of mass energy of the colliders used to place the constraints), any
corrections should be of order $m/E$ or smaller and hence we do not expect our conclusions to change if one were to include a
small mass in their analysis.}. The simplest model studied here has $B_\infty=1$, for which we find
$\mathcal{M}\gsim1\textrm{eV}$, where
we have taken $H_0^2\mpl^2\sim 10^{-12}$ eV$^4$. One can see that the laboratory constraints are eleven orders-of-magnitude
stronger than the most stringent bounds we obtain here. This is because the local scalar charge is due entirely to the
cosmological
motion of the field, which gives rise to an effective three-point interaction with coupling constant $g\equiv
\mpl^2B_\infty\dot{\phi}_\infty/\Lambda^2$. This appears in the vertices of the tree-level 2-2 scattering amplitude and so the
classical force constrains a combination of $\dot{\phi}\inff$, $\ddot{\phi}_\infty$ and $\Lambda$. At loop-level, there are also
four-point vertices that depend on $\mathcal{M}^4\sim\mpl^2\Lambda^2$ alone. For this reason, experiments that probe the quantum
effects can probe $\Lambda$ directly whereas solar system tests are limited by the background cosmology, which aids in satisfying
the constraints.

\section{Discussion}
\label{sec:disc}
In this work we have shown how the cosmological evolution of the scalar field can have implications on smaller scales if there is
a disformal coupling to matter. Having investigated the nature of the local modifications of gravity, we are in a position to
classify how well the theory is screened in different environments and so before concluding, in this section we pause to discuss
the prospects for constraining these theories further. We also discuss the implications of the results found here for other
theories of modified gravity that contain disformal couplings in the decoupling limit. Before doing so, we discuss the
implications of our new constraints for the cosmology of disformally coupled metrics.

\subsection{Implications for Disformal Cosmology}

The models we investigated in section \ref{sec:cosmo} all had $\epsilon_\Lambda\sim\mathcal{O}(1)$ or more, but the constraint
from the Cassini bound implies $\epsilon_\Lambda\lsim 10^{-12}$. Numerically, we find that models with values of
$\epsilon_\Lambda$ this small are indistinguishable from the equivalent quintessence model, at least at the background level. The
Cassini bound comes with a caveat: In section \ref{sec:cosmo} we found that large values of $\epsilon_\Lambda$ can delay the
onset of dark energy domination. One can then imagine a model where $\epsilon_\Lambda$ is large enough such that there is zero
scalar charge locally. Such a model would be incompatible with the constraints of \cite{Kaloper:2003yf,Brax:2014vva,Brax:2014zba}, which are
independent of the cosmological dynamics. In this case the expansion history would be indistinguishable from $\Lambda$CDM. One possible class 
of exceptions to this are more complicated models where $B(\phi)$ and $\gamma(\phi)$ vary greatly during the evolution of the universe. In this case, 
the Cassini bound would constrain their values today and deviations in the past could drive differences in the expansion history at early times.

\subsection{Prospects for Future Constraints}

Below, we briefly comment on the prospects for obtaining further constraints in different physical systems.

\begin{itemize}
 
\item {\bf Solar system}: These theories satisfy the equivalence principle, which has prevented us from utilising some of the
solar system experiments to place constraints. Furthermore, since the force range is of order $H_0^{-1}$, the fifth-force is
inverse-square with $\Delta G=2Q^2G$. Effects that result from deviations from the inverse-square law such as the precession of
the Earth-Moon perigee or the precession of Mercury are therefore absent. One can circumvent this argument if one reduces the
force range, however, this means that the mass of the field is greater than Hubble and so the cosmological field should rest at
its minimum, in which case $Q$ is identically zero. There is one loop hole in this argument, namely that it relies on our
intuition from quintessence models, which may not apply to disformally coupled theories. Indeed, in section \ref{sec:cosmo} we
found that a large disformal coupling (small $\Lambda$) can delay the time at which the field begins to roll. A full study of this
effect requires one to find the best-fitting cosmological model, which is beyond the scope of this work. Here, we note that the
constraints of \cite{Kaloper:2003yf,Brax:2014vva,Brax:2014zba} are robust to this effect and so we expect any model that circumvents the Cassini
bound by exploiting this loop hole to be ruled out by their results\footnote{In fact, there is a further caveat. Since their
results assume a massless scalar one may be able to find a model where the laboratory tests are satisfied and the field has a
large mass and disformal coupling such that it has only begun to roll in the recent past.}. An important class of exceptions to
the statements above are models that do not have a minimum such as exponentials and inverse-power laws. These would still be
rolling today and so may show effects in the solar system.

\item {\bf Astrophysical systems}: Astrophysical tests have provided the strongest constraints on chameleon models to
date \cite{Chang:2010xh,Jain:2011ji,Davis:2011qf,Jain:2012tn,Vikram:2013uba,Sakstein:2013pda,Vikram:2014uza}. Several of these
tests are due to the modified stellar structure and pulsation frequencies. This is because some stars can be very unscreened
provided that they reside in dwarf galaxies in cosmological voids. In disformal theories, we do not have this environmental
dependence and the scalar charge $Q$ is universal. For this reason, stellar tests will likely be less accurate than solar system
probes due to a combination of a small signal and large astrophysical uncertainties. Similarly, all of the galactic tests rely on
equivalence principle violations, which are absent in these theories.

\item {\bf Background Cosmology}: We have seen that this model is indistinguishable from the equivalent quintessence model when we
impose the constraint from the Cassini probe. Therefore, the background expansion is identical. The one caveat to this is the
fact that $G\eff$ can vary in the past and so probes of the expansion history such as type Ia supernovae light curves may vary
with redshift leading to different predictions for the Luminosity distance. Furthermore, we have not discussed the issue of the
initial conditions or the evolution during the radiation dominated era. Chameleon models experience kicks due to the decoupling
of non-relativistic species in the radiation dominated epoch \cite{Brax:2004qh}. It would be interesting to calculate what effect,
if any, this has on the evolution of disformally coupled theories.

\item {\bf Linear and non-linear cosmological probes}: We have not studied the relativistic perturbations about the FRW solution
in this paper, however, this was studied by \cite{Zumalacarregui:2012us} who found large deviations in the CDM and
CMB power spectra from the $\Lambda$CDM predictions. The authors used a different model to the one presented here, which, unlike
ours, was not designed to give screening in the solar system at late times and it is possible that different models may
predict smaller deviations. Importantly, the requirement of screening in the solar system applies only at present times.
Chameleon-like theories and those that screen using the Vainshtein mechanism are screened at all times due to non-linear effects
but this is not necessarily the case for disformal models. Indeed, most models of quintessence have larger time-derivatives in
the past and so it is likely that these theories are more unscreened at early times, which could lead to large deviations from
$\Lambda$CDM in the linear and non-linear regime. This may be the most efficient method of probing model's whose mass is larger
than Hubble since they have zero scalar charge in the solar system today.

\item {\bf Laboratory tests}: These typically probe length scales between a few microns to $10^{14}$ m \cite{Adelberger:2003zx}.
As mentioned above, any theory with a Compton wavelength in this range has a mass far greater than the Hubble parameter and has
likely reached the minimum of its potential by the current epoch, in which case the theory is indistinguishable from GR on small
scales. Again, models that do not have a minimum may circumvent this argument. This argument applies to tree-level effects
only. Recently, \cite{Kaloper:2003yf,Brax:2014vva,Brax:2014zba} have shown that loop effects
from the coupling $T\mmm^{\mu\nu}\partial_\mu\phi\partial_\nu\phi$ can be important and, indeed, here we have found that their
constraints are stronger than ours by several orders-of-magnitude. This is therefore a promising avenue and future studies
concentrated here may yield even stronger bounds.

\item {\bf Strongly gravitating systems:} In this work we have only considered non-relativistic pressureless sources. One may then
wonder whether the pressure coupling in equation (\ref{eq:sfeom}) could lead to observable consequences. There are two potential
places where this may be non negligible: The early universe and compact objects. The first has been studied by
\cite{vandeBruck:2012vq,vandeBruck:2013yxa} who study linear perturbations during the radiation epoch and the spectral distortion
of the CMB. Interestingly, they point out that the equation of state in the Einstein frame is modified when the pressure is
non-negligible. As for the second, one may wonder whether the structure of Neutron stars may be significantly altered. One can
argue that this is generically not the case. The equation of motion for a fully
relativistic, static star is 
\begin{equation}\label{eq:NS}
 \left[\left(1-\frac{B{\phi^\prime}^2}{\Lambda^2}\right)-\frac{8\pi GB^2 P}{\Lambda^2}\right]\nabla_r\nabla_r\phi=8\pi
\gamma(\phi)G\frac{B^2}{\Lambda^2}T\mmm^\nm\partial_\mu\phi\partial_\nu\phi+\left(1-\frac{B{\phi^\prime}^2}{\Lambda^2}
\right)V_{,\phi}.
\end{equation}
First, note that the only source is the derivative of the potential, $V(\phi)_{,\phi}\sim m^2\phi$ where $m$ is the field's
mass. When
$mR\ll1$---R is the star's radius---as we expect for cosmological scalars we can ignore this and the space-time is asymptotically
Minkowski. In this case, $\phi(r)=\textrm{constant}$ is a consistent solution of the equation and the star's structure is
identical to the GR prediction. When we include the expansion of the universe the right hand side of (\ref{eq:NS}) contains a
time-varying source that is the analogue of $Q$ in the non-relativistic limit and an additional pressure coupling:
\begin{equation}
 \left[\left(1-\frac{B{\phi^\prime}^2}{\Lambda^2}\right)-\frac{8\pi GB^2 P}{\Lambda^2}\right]\nabla_r\nabla_r\phi=8\pi
\gamma(\phi)G\frac{B^2}{\Lambda^2}P{\phi^\prime}^2+8\pi Q\rho.
\end{equation}
We have already seen that $Q$ is constrained to be of order $10^{-12}$ or less by solar system bounds and one would not expect
this to differ drastically in fully relativistic settings because $Q$ is sourced by the relativistic FRW background in both
scenarios. Whether or not this source is enhanced or suppressed by relativistic effects depends on whether the factor multiplying
the Laplacian differs greatly from $1$. Spherical symmetry demands that $\phi^\prime(r\rightarrow0)=0$ and so we expect
$\phi^\prime\ll\Lambda$ near the centre of the star and hence the source is vastly suppressed. One would therefore not expect
neutron stars to give stronger bounds than non-relativistic objects, especially when one accounts for degeneracies with the
equation of state and the fact that one must perform less precise measurements outside the solar system. This conclusion may
change when time-dependence is important, for example in rapidly rotating neutron stars.

\item {\bf Binary Pulsars}: Scalar-tensor theories of gravity predict mono-, di- and quadrupolar radiation from binary pulsar
systems not
predicted in GR \cite{Damour:1992we}. This leads to weak constraints for chameleon models \cite{Brax:2013uh} and predictions that
are currently unobservable for the cubic galileon \cite{deRham:2012fg}. TeVeS \cite{Bekenstein:2004ne} contains a scalar field
disformally coupled to matter and \cite{Freire:2012mg} argue that the functional forms (see also
\cite{Damour:1995kt,Damour:1996ke}) of the formulae for the emission of scalar radiation are unchanged in these theories, only
the numerical coefficients change due to the altered composition of the stars. The diploar emission
vanishes when the scalar charges are equal, as is the case for disformal theories. The mono- and quadrupolar emission is
proportional to the product of the scalar charges, which we have seen above is constrained by solar system tests to be $\lsim
10^{-24}$ and so we can conclude that the emission of scalar radiation is negligible compared with the gravitational radiation.
It is therefore unlikely that binary pulsars are a good probe of these theories. 

\end{itemize}

\subsection{Implications for Other Theories of Modified Gravity}

In this work we have focused on the specific class of theories (\ref{eq:act}), but, as argued in section \ref{sec:dec}, we could
just as easily have started from the decoupling limit
\begin{equation}\label{eq:declim}
 \frac{\mathcal{L}}{\sqrt{-g}}=
-\frac{1}{4}h^\nm\left(\mathcal{E}h\right)_\nm+\cdots+T\mmm\ln A(\phi)+
\frac{1}{2{\Lambda}^2}T^\nm\partial_\mu\phi\partial_\nu\phi,
\end{equation}
where $(\mathcal{E}h)_\nm$ is the Lichnerowicz operator and the dots represent arbitrary kinetic and potential terms. In this
case, an effective scalar charge $Q$ would be obtained by expanding $\phi=\phii+\vpp$ so that $\nabla^2\vpp+\cdots=8\pi
(\alpha+Q)G\rho$, where we have separated the disformal and conformal contribution to the equation of motion. Equation
(\ref{eq:declim}) does not have a unique covariantisation. Indeed, a theory with disformally coupled metric is one, especially if
we take the kinetic term to be canonical but the results of this work are more general than this. Indeed, if one allows for
arbitrary galileon terms \cite{Nicolis:2008in} then an equally valid covariantisation is the covariant galileon branch
\cite{Deffayet:2009wt} of the Horndeski theory \cite{Horndeski:1974wa}. In this case, the covariantisation of the disformal
coupling is included in the choice of the function $G(\phi)$ ($\mathcal{L}/\sqrt{-g}\supset
G_5(\phi)G_{\nm}\nabla^\mu\nabla^\nu\phi$) \cite{Appleby:2011aa}. These two theories are not equivalent because one cannot
define an Einstein frame when such a term is present \cite{Bettoni:2013diz}. Other equally valid covariantisations are massive
gravity \cite{deRham:2014zqa,Ondo:2013wka} and massive bi-gravity \cite{Hassan:2011zd}, which both contain disformal interactions
in the decoupling limit.
%

The disformal coupling is often neglected when examining the behaviour of these theories in the solar system because it vanishes
in the static limit when taking the space-time to be Minkowski. What we have shown here is that when one considers the more
realistic scenario where the field has a time-dependent VEV, this coupling can contribute to the equation of motion, even in the
static limit. Galileon theories have $\ln A=\alpha$, where $\alpha$ is a constant that is typically taken to be $\mathcal{O}(1)$.
Including the disformal coupling, one has an effective coupling
\begin{equation}\label{eq:alphaeff}
 \alpha\eff=\alpha+\frac{{\ddot{\phi}_\infty}}{\Lambda^2}.
\end{equation}
Now in the theories described so far, we have vanishing time-derivatives (slow-roll) at late times but galileon theories
self-accelerate
and do not have this property. Indeed, \cite{DeFelice:2010pv} find a tracker solution for the cubic galileon at late times such
that $\dot{\phi}_\infty H\equiv \xi H_0$ is constant. One can see qualitatively how the disformal
contribution to the effective coupling can play a role in the small scale physics. For simplicity, let us work with the cubic
galileon, in which case the Vainshtein radius is \cite{Khoury:2013tda}
\begin{equation}
 r_{\rm V}=\left(\alpha r_{\rm S}L^2\right)^{\frac{1}{3}},
\end{equation}
where $r_{\rm S}$ is the Schwarzchild radius. Typically, $\alpha$ is assumed to be or order unity, in which case LLR constraints
on deviations from the inverse-square law give
$L\gsim 150$ Mpc \cite{Afshordi:2008rd}. Including the disformal factor, this is modified to 
\begin{equation}
 \sqrt{\alpha+\frac{\ddot{\phi}_\infty}{\Lambda^2}}L>150\textrm{Mpc}.
\end{equation}
One can see that the cosmological dynamics can change the bounds on the Vainshtein radius of the Earth. Furthermore, since
$\Lambda_3=(\mpl L^{-2})$, where the cubic galileon kinetic term is $-\mpl\Lambda_3^{-3}(\partial\phi)^2\Box\phi$, one can see
that solar system tests no longer constrain $\Lambda_3$ alone but rather a region in the $\Lambda_3$--$\Lambda$ plane. This
new effect certainly merits further investigation.

\section{Conclusions}
\label{sec:concs}
This paper has studied theories of gravity where a scalar field is disformally coupled to matter. Studies of theories with
this coupling often neglect it on small scales because the non-relativistic limit is static and there are therefore no
contributions to the equation of motion. This is only correct in space-times that are non-relativistic perturbations about
Minkowski space. What we have shown here is that by considering perturbations about an expanding FRW space-time---which is the
physically relevant scenario---this coupling can result in unscreened fifth-forces on all scales whose magnitude depends on the
model parameters and the first and second time-derivatives of the homogeneous component of the scalar field\footnote{Note that this means that the 
local scalar charge depends on the time-derivatives of the cosmological field only. The local time-derivatives are not important at Newtonian order 
but may be relevant for relativistic systems such as rotating neutron stars.}. The local scalar charge vanishes when these time-derivatives are zero 
and we hence argue that any theory where the cosmological scalar is rolling down a potential should be
able to screen linearly (i.e. the modifications to gravity are absent on all scales) at late times.

This was investigated using the simplest model expected to exhibit these features, namely a constant
disformal factor and a quadratic scalar potential.The cosmological Friedmann-scalar field system was solved numerically and it was
found
that the behaviour of the system is similar to the equivalent quintessence system but that larger disformal couplings can delay
the time at which the field begins to roll. 

Next, local tests of gravity---the Cassini probe and lunar laser ranging---were used to place new bounds on the model
parameters. These theories satisfy the equivalence principle,
which prevented us from utilising other solar system tests. The field's mass is expected to be of order Hubble today
otherwise the scalar charge is identically zero. For this reason, laboratory probes are not sensitive to these theories. We find
that $\Lambda/H_0>5.6\times10^5$, or, in the language of
\cite{Kaloper:2003yf,Brax:2014vva,Brax:2014zba}, $\mathcal{M}\gsim\mathcal{O}(\textrm{eV})$. Once
these bounds are imposed, the background cosmology is indistinguishable from the equivalent quintessence model. Our
bounds are not competitive with those found by \cite{Kaloper:2003yf,Brax:2014vva,Brax:2014zba} by several
orders-of-magnitude. 

We discussed the future prospects for testing these theories in other systems. Laboratory tests of the quantum effects
seem promising because they do not depend on the cosmology. Furthermore, deviations on linear and non-linear scales may be useful
because these theories are more unscreened in the past.

Finally, our results have implications for other theories of modified gravity. Any theory that contains
the disformal coupling should exhibit the features investigated here and theories that screen using the
Vainshtein mechanism may have Vainshtein radii vastly different from the n\"{a}ive prediction because large cosmological
time-derivatives can change the effective coupling to matter from $\mathcal{O}(1)$.

Several times we have stated the caveats to the bounds obtained here. Firstly, they only apply to models whose cosmological masses
are of order Hubble. Theories with larger masses are expected to lie at the minimum of their potentials today and hence have zero
scalar charge. The exception to this are models with no minima such as exponentials or inverse-power laws. Secondly, larger
disformal couplings may evade the bounds because the field has not yet begun to roll, although these models are ruled out
independently by \cite{Kaloper:2003yf,Brax:2014vva,Brax:2014zba}\footnote{Again, there is the caveat that the authors assume a massless scalar
in their calculations.}. Clearly there are many models left to explore and here we have focused on the simplest one in
order to elucidate the screening properties of the theory. Finding a viable model with falsifiable predictions is of paramount
importance and so future studies should certainly begin here.

\section*{Acknowledgements}

I would like to thank Dario Bettoni, Adam Solomon and Miguel Zumalac\'{a}rregui for their collaboration and comments during the
early stages of this work. I am incredibly grateful to Niayesh Afshordi for several enlightening and stimulating discussions and a
careful reading of the manuscript. I am also grateful to Tomi Koivisto and David F. Mota for their comments on an early draft. This work has 
benefited from discussions with Alexandre Barreira, Philippe Brax, Clare Burrage, Anne-Christine Davis, Kurt Hinterbichler, Kazuya Koyama, Baojiu Li, 
Claudia de Rahm and Mark Trodden.

\bibliographystyle{JHEP}
\bibliography{ref}

\providecommand{\href}[2]{#2}\begingroup\raggedright\begin{thebibliography}{10}

\bibitem{Clifton:2011jh}
T.~Clifton, P.~G. Ferreira, A.~Padilla, and C.~Skordis, {\it {Modified Gravity
  and Cosmology}},  {\em Phys.Rept.} {\bf 513} (2012) 1--189,
  [\href{http://arxiv.org/abs/1106.2476}{{\tt arXiv:1106.2476}}].

\bibitem{Jain:2010ka}
B.~Jain and J.~Khoury, {\it {Cosmological Tests of Gravity}},  {\em Annals
  Phys.} {\bf 325} (2010) 1479--1516,
  [\href{http://arxiv.org/abs/1004.3294}{{\tt arXiv:1004.3294}}].

\bibitem{Joyce:2014kja}
A.~Joyce, B.~Jain, J.~Khoury, and M.~Trodden, {\it {Beyond the Cosmological
  Standard Model}},  \href{http://arxiv.org/abs/1407.0059}{{\tt
  arXiv:1407.0059}}.

\bibitem{Khoury:2003aq}
J.~Khoury and A.~Weltman, {\it {Chameleon fields: Awaiting surprises for tests
  of gravity in space}},  {\em Phys.Rev.Lett.} {\bf 93} (2004) 171104,
  [\href{http://arxiv.org/abs/astro-ph/0309300}{{\tt astro-ph/0309300}}].

\bibitem{Khoury:2003rn}
J.~Khoury and A.~Weltman, {\it {Chameleon Cosmology}},  {\em Phys. Rev.} {\bf
  D69} (2004) 044026, [\href{http://arxiv.org/abs/astro-ph/0309411}{{\tt
  astro-ph/0309411}}].

\bibitem{Hinterbichler:2010es}
K.~Hinterbichler and J.~Khoury, {\it {Symmetron Fields: Screening Long-Range
  Forces Through Local Symmetry Restoration}},  {\em Phys. Rev. Lett.} {\bf
  104} (2010) 231301, [\href{http://arxiv.org/abs/1001.4525}{{\tt
  arXiv:1001.4525}}].

\bibitem{Brax:2010gi}
P.~Brax, C.~van~de Bruck, A.-C. Davis, and D.~Shaw, {\it {The Dilaton and
  Modified Gravity}},  {\em Phys. Rev.} {\bf D82} (2010) 063519,
  [\href{http://arxiv.org/abs/1005.3735}{{\tt arXiv:1005.3735}}].

\bibitem{Vainshtein:1972sx}
A.~Vainshtein, {\it {To the problem of nonvanishing gravitation mass}},  {\em
  Phys.Lett.} {\bf B39} (1972) 393--394.

\bibitem{Nicolis:2008in}
A.~Nicolis, R.~Rattazzi, and E.~Trincherini, {\it {The Galileon as a local
  modification of gravity}},  {\em Phys.Rev.} {\bf D79} (2009) 064036,
  [\href{http://arxiv.org/abs/0811.2197}{{\tt arXiv:0811.2197}}].

\bibitem{Hinterbichler:2011tt}
K.~Hinterbichler, {\it {Theoretical Aspects of Massive Gravity}},  {\em
  Rev.Mod.Phys.} {\bf 84} (2012) 671--710,
  [\href{http://arxiv.org/abs/1105.3735}{{\tt arXiv:1105.3735}}].

\bibitem{deRham:2014zqa}
C.~de~Rham, {\it {Massive Gravity}},
  \href{http://arxiv.org/abs/1401.4173}{{\tt arXiv:1401.4173}}.

\bibitem{1992mgm..conf..905B}
J.~D. {Bekenstein}, {\it {New gravitational theories as alternatives to dark
  matter.}},  in {\em Marcel Grossmann Meeting on General Relativity}
  (F.~{Sat{\= o}} and T.~{Nakamura}, eds.), p.~905, 1992.

\bibitem{Bekenstein:1992pj}
J.~D. Bekenstein, {\it {The Relation between physical and gravitational
  geometry}},  {\em Phys.Rev.} {\bf D48} (1993) 3641--3647,
  [\href{http://arxiv.org/abs/gr-qc/9211017}{{\tt gr-qc/9211017}}].

\bibitem{Zumalacarregui:2013pma}
M.~Zumalacárregui and J.~García-Bellido, {\it {Transforming gravity: from
  derivative couplings to matter to second-order scalar-tensor theories beyond
  the Horndeski Lagrangian}},  {\em Phys.Rev.} {\bf D89} (2014) 064046,
  [\href{http://arxiv.org/abs/1308.4685}{{\tt arXiv:1308.4685}}].

\bibitem{Koivisto:2008ak}
T.~S. Koivisto, {\it {Disformal quintessence}},
  \href{http://arxiv.org/abs/0811.1957}{{\tt arXiv:0811.1957}}.

\bibitem{Zumalacarregui:2010wj}
M.~Zumalacarregui, T.~Koivisto, D.~Mota, and P.~Ruiz-Lapuente, {\it {Disformal
  Scalar Fields and the Dark Sector of the Universe}},  {\em JCAP} {\bf 1005}
  (2010) 038, [\href{http://arxiv.org/abs/1004.2684}{{\tt arXiv:1004.2684}}].

\bibitem{Kaloper:2003yf}
N.~Kaloper, {\it {Disformal inflation}},  {\em Phys.Lett.} {\bf B583} (2004)
  1--13, [\href{http://arxiv.org/abs/hep-ph/0312002}{{\tt hep-ph/0312002}}].

\bibitem{Koivisto:2012za}
T.~S. Koivisto, D.~F. Mota, and M.~Zumalacarregui, {\it {Screening
  Modifications of Gravity through Disformally Coupled Fields}},  {\em
  Phys.Rev.Lett.} {\bf 109} (2012) 241102,
  [\href{http://arxiv.org/abs/1205.3167}{{\tt arXiv:1205.3167}}].

\bibitem{Zumalacarregui:2012us}
M.~Zumalacarregui, T.~S. Koivisto, and D.~F. Mota, {\it {DBI Galileons in the
  Einstein Frame: Local Gravity and Cosmology}},  {\em Phys.Rev.} {\bf D87}
  (2013) 083010, [\href{http://arxiv.org/abs/1210.8016}{{\tt
  arXiv:1210.8016}}].

\bibitem{vandeBruck:2012vq}
C.~van~de Bruck and G.~Sculthorpe, {\it {Modified Gravity and the Radiation
  Dominated Epoch}},  {\em Phys.Rev.} {\bf D87} (2013), no.~4 044004,
  [\href{http://arxiv.org/abs/1210.2168}{{\tt arXiv:1210.2168}}].

\bibitem{Brax:2013nsa}
P.~Brax, C.~Burrage, A.-C. Davis, and G.~Gubitosi, {\it {Cosmological Tests of
  the Disformal Coupling to Radiation}},  {\em JCAP} {\bf 1311} (2013) 001,
  [\href{http://arxiv.org/abs/1306.4168}{{\tt arXiv:1306.4168}}].

\bibitem{vandeBruck:2013yxa}
C.~van~de Bruck, J.~Morrice, and S.~Vu, {\it {Constraints on Nonconformal
  Couplings from the Properties of the Cosmic Microwave Background Radiation}},
   {\em Phys.Rev.Lett.} {\bf 111} (2013) 161302,
  [\href{http://arxiv.org/abs/1303.1773}{{\tt arXiv:1303.1773}}].

\bibitem{Brax:2014vva}
P.~Brax and C.~Burrage, {\it {Constraining Disformally Coupled Scalar Fields}},
   \href{http://arxiv.org/abs/1407.1861}{{\tt arXiv:1407.1861}}.

\bibitem{Brax:2014zba}
P.~Brax and C.~Burrage, {\it {Explaining the Proton Radius Puzzle with
  Disformal Scalars}},  \href{http://arxiv.org/abs/1407.2376}{{\tt
  arXiv:1407.2376}}.

\bibitem{Noller:2012sv}
J.~Noller, {\it {Derivative Chameleons}},  {\em JCAP} {\bf 1207} (2012) 013,
  [\href{http://arxiv.org/abs/1203.6639}{{\tt arXiv:1203.6639}}].

\bibitem{Damour:1994zq}
T.~Damour and A.~M. Polyakov, {\it {The String dilaton and a least coupling
  principle}},  {\em Nucl. Phys.} {\bf B423} (1994) 532--558,
  [\href{http://arxiv.org/abs/hep-th/9401069}{{\tt hep-th/9401069}}].

\bibitem{Hui:2009kc}
L.~Hui, A.~Nicolis, and C.~Stubbs, {\it {Equivalence Principle Implications of
  Modified Gravity Models}},  {\em Phys.Rev.} {\bf D80} (2009) 104002,
  [\href{http://arxiv.org/abs/0905.2966}{{\tt arXiv:0905.2966}}].

\bibitem{Koivisto:2013fta}
T.~Koivisto, D.~Wills, and I.~Zavala, {\it {Dark D-brane Cosmology}},
  \href{http://arxiv.org/abs/1312.2597}{{\tt arXiv:1312.2597}}.

\bibitem{Copeland:2006wr}
E.~J. Copeland, M.~Sami, and S.~Tsujikawa, {\it {Dynamics of dark energy}},
  {\em Int.J.Mod.Phys.} {\bf D15} (2006) 1753--1936,
  [\href{http://arxiv.org/abs/hep-th/0603057}{{\tt hep-th/0603057}}].

\bibitem{Barreira:2013xea}
A.~Barreira, B.~Li, C.~Baugh, and S.~Pascoli, {\it {Spherical collapse in
  Galileon gravity: fifth force solutions, halo mass function and halo bias}},
  \href{http://arxiv.org/abs/1308.3699}{{\tt arXiv:1308.3699}}.

\bibitem{deRham:2010kj}
C.~de~Rham, G.~Gabadadze, and A.~J. Tolley, {\it {Resummation of Massive
  Gravity}},  {\em Phys.Rev.Lett.} {\bf 106} (2011) 231101,
  [\href{http://arxiv.org/abs/1011.1232}{{\tt arXiv:1011.1232}}].

\bibitem{Appleby:2011aa}
S.~Appleby and E.~V. Linder, {\it {The Paths of Gravity in Galileon
  Cosmology}},  {\em JCAP} {\bf 1203} (2012) 043,
  [\href{http://arxiv.org/abs/1112.1981}{{\tt arXiv:1112.1981}}].

\bibitem{Bertotti:2003rm}
B.~Bertotti, L.~Iess, and P.~Tortora, {\it {A test of general relativity using
  radio links with the Cassini spacecraft}},  {\em Nature} {\bf 425} (2003)
  374.

\bibitem{Turyshev:2006gm}
S.~G. Turyshev and J.~G. Williams, {\it {Space-based tests of gravity with
  laser ranging}},  {\em Int.J.Mod.Phys.} {\bf D16} (2007) 2165--2179,
  [\href{http://arxiv.org/abs/gr-qc/0611095}{{\tt gr-qc/0611095}}].

\bibitem{Mota:2006fz}
D.~F. Mota and D.~J. Shaw, {\it {Evading equivalence principle violations,
  astrophysical and cosmological constraints in scalar field theories with a
  strong coupling to matter}},  {\em Phys. Rev.} {\bf D75} (2007) 063501,
  [\href{http://arxiv.org/abs/hep-ph/0608078}{{\tt hep-ph/0608078}}].

\bibitem{Brax:2008hh}
P.~Brax, C.~van~de Bruck, A.-C. Davis, and D.~J. Shaw, {\it {f(R) Gravity and
  Chameleon Theories}},  {\em Phys.Rev.} {\bf D78} (2008) 104021,
  [\href{http://arxiv.org/abs/0806.3415}{{\tt arXiv:0806.3415}}].

\bibitem{Brax:2007vm}
P.~Brax, C.~van~de Bruck, A.-C. Davis, D.~F. Mota, and D.~J. Shaw, {\it
  {Detecting chameleons through Casimir force measurements}},  {\em Phys.Rev.}
  {\bf D76} (2007) 124034, [\href{http://arxiv.org/abs/0709.2075}{{\tt
  arXiv:0709.2075}}].

\bibitem{Brax:2009bk}
P.~Brax, C.~van~de Bruck, A.-C. Davis, and D.~Shaw, {\it {Laboratory Tests of
  Chameleon Models}},  \href{http://arxiv.org/abs/0911.1086}{{\tt
  arXiv:0911.1086}}.

\bibitem{Brax:2011sv}
P.~Brax, C.~Burrage, and A.-C. Davis, {\it {Laboratory Tests of the Galileon}},
   {\em JCAP} {\bf 1109} (2011) 020,
  [\href{http://arxiv.org/abs/1106.1573}{{\tt arXiv:1106.1573}}].

\bibitem{Kapner:2006si}
D.~Kapner, T.~Cook, E.~Adelberger, J.~Gundlach, B.~R. Heckel, et~al., {\it
  {Tests of the gravitational inverse-square law below the dark-energy length
  scale}},  {\em Phys.Rev.Lett.} {\bf 98} (2007) 021101,
  [\href{http://arxiv.org/abs/hep-ph/0611184}{{\tt hep-ph/0611184}}].

\bibitem{Hoskins:1985tn}
J.~Hoskins, R.~Newman, R.~Spero, and J.~Schultz, {\it {Experimental tests of
  the gravitational inverse square law for mass separations from 2-cm to
  105-cm}},  {\em Phys.Rev.} {\bf D32} (1985) 3084--3095.

\bibitem{Lambrecht:2011qm}
A.~Lambrecht and S.~Reynaud, {\it {Casimir and short-range gravity tests}},
  \href{http://arxiv.org/abs/1106.3848}{{\tt arXiv:1106.3848}}.

\bibitem{Chang:2010xh}
P.~Chang and L.~Hui, {\it {Stellar Structure and Tests of Modified Gravity}},
  \href{http://arxiv.org/abs/1011.4107}{{\tt arXiv:1011.4107}}.

\bibitem{Jain:2011ji}
B.~Jain and J.~VanderPlas, {\it {Tests of Modified Gravity with Dwarf
  Galaxies}},  {\em JCAP} {\bf 1110} (2011) 032,
  [\href{http://arxiv.org/abs/1106.0065}{{\tt arXiv:1106.0065}}].

\bibitem{Davis:2011qf}
A.-C. Davis, E.~A. Lim, J.~Sakstein, and D.~Shaw, {\it {Modified Gravity Makes
  Galaxies Brighter}},  {\em Phys.Rev.} {\bf D85} (2012) 123006,
  [\href{http://arxiv.org/abs/1102.5278}{{\tt arXiv:1102.5278}}].

\bibitem{Jain:2012tn}
B.~Jain, V.~Vikram, and J.~Sakstein, {\it {Astrophysical Tests of Modified
  Gravity: Constraints from Distance Indicators in the Nearby Universe}},  {\em
  Astrophys.J.} {\bf 779} (2013) 39,
  [\href{http://arxiv.org/abs/1204.6044}{{\tt arXiv:1204.6044}}].

\bibitem{Vikram:2013uba}
V.~Vikram, A.~Cabre, B.~Jain, and J.~VanderPlas, {\it {Astrophysical Tests of
  Modified Gravity: the Morphology and Kinematics of Dwarf Galaxies}},
  \href{http://arxiv.org/abs/1303.0295}{{\tt arXiv:1303.0295}}.

\bibitem{Sakstein:2013pda}
J.~Sakstein, {\it {Stellar Oscillations in Modified Gravity}},  {\em Phys.Rev.}
  {\bf D88} (2013) 124013, [\href{http://arxiv.org/abs/1309.0495}{{\tt
  arXiv:1309.0495}}].

\bibitem{Vikram:2014uza}
V.~Vikram, J.~Sakstein, C.~Davis, and A.~Neil, {\it {Astrophysical Tests of
  Modified Gravity: Stellar and Gaseous Rotation Curves in Dwarf Galaxies}},
  \href{http://arxiv.org/abs/1407.6044}{{\tt arXiv:1407.6044}}.

\bibitem{Brax:2004qh}
P.~Brax, C.~van~de Bruck, A.-C. Davis, J.~Khoury, and A.~Weltman, {\it
  {Detecting dark energy in orbit: The cosmological chameleon}},  {\em Phys.
  Rev.} {\bf D70} (2004) 123518,
  [\href{http://arxiv.org/abs/astro-ph/0408415}{{\tt astro-ph/0408415}}].

\bibitem{Adelberger:2003zx}
E.~Adelberger, B.~R. Heckel, and A.~Nelson, {\it {Tests of the gravitational
  inverse square law}},  {\em Ann.Rev.Nucl.Part.Sci.} {\bf 53} (2003) 77--121,
  [\href{http://arxiv.org/abs/hep-ph/0307284}{{\tt hep-ph/0307284}}].

\bibitem{Damour:1992we}
T.~Damour and G.~Esposito-Farese, {\it {Tensor multiscalar theories of
  gravitation}},  {\em Class.Quant.Grav.} {\bf 9} (1992) 2093--2176.

\bibitem{Brax:2013uh}
P.~Brax, A.-C. Davis, and J.~Sakstein, {\it {Pulsar Constraints on Screened
  Modified Gravity}},  \href{http://arxiv.org/abs/1301.5587}{{\tt
  arXiv:1301.5587}}.

\bibitem{deRham:2012fg}
C.~de~Rham, A.~Matas, and A.~J. Tolley, {\it {Galileon Radiation from Binary
  Systems}},  {\em Phys.Rev.} {\bf D87} (2013), no.~6 064024,
  [\href{http://arxiv.org/abs/1212.5212}{{\tt arXiv:1212.5212}}].

\bibitem{Bekenstein:2004ne}
J.~D. Bekenstein, {\it {Relativistic gravitation theory for the MOND
  paradigm}},  {\em Phys.Rev.} {\bf D70} (2004) 083509,
  [\href{http://arxiv.org/abs/astro-ph/0403694}{{\tt astro-ph/0403694}}].

\bibitem{Freire:2012mg}
P.~C. Freire, N.~Wex, G.~Esposito-Farese, J.~P. Verbiest, M.~Bailes, et~al.,
  {\it {The relativistic pulsar-white dwarf binary PSR J1738+0333 II. The most
  stringent test of scalar-tensor gravity}},  {\em Mon.Not.Roy.Astron.Soc.}
  {\bf 423} (2012) 3328, [\href{http://arxiv.org/abs/1205.1450}{{\tt
  arXiv:1205.1450}}].

\bibitem{Damour:1995kt}
T.~Damour and G.~Esposito-Farese, {\it {Testing gravity to second postNewtonian
  order: A Field theory approach}},  {\em Phys.Rev.} {\bf D53} (1996)
  5541--5578, [\href{http://arxiv.org/abs/gr-qc/9506063}{{\tt gr-qc/9506063}}].

\bibitem{Damour:1996ke}
T.~Damour and G.~Esposito-Farese, {\it {Tensor - scalar gravity and binary
  pulsar experiments}},  {\em Phys.Rev.} {\bf D54} (1996) 1474--1491,
  [\href{http://arxiv.org/abs/gr-qc/9602056}{{\tt gr-qc/9602056}}].

\bibitem{Deffayet:2009wt}
C.~Deffayet, G.~Esposito-Farese, and A.~Vikman, {\it {Covariant Galileon}},
  {\em Phys.Rev.} {\bf D79} (2009) 084003,
  [\href{http://arxiv.org/abs/0901.1314}{{\tt arXiv:0901.1314}}].

\bibitem{Horndeski:1974wa}
G.~W. Horndeski, {\it {Second-order scalar-tensor field equations in a
  four-dimensional space}},  {\em Int.J.Theor.Phys.} {\bf 10} (1974) 363--384.

\bibitem{Bettoni:2013diz}
D.~Bettoni and S.~Liberati, {\it {Disformal invariance of second order
  scalar-tensor theories: Framing the Horndeski action}},  {\em Phys.Rev.} {\bf
  D88} (2013), no.~8 084020, [\href{http://arxiv.org/abs/1306.6724}{{\tt
  arXiv:1306.6724}}].

\bibitem{Ondo:2013wka}
N.~A. Ondo and A.~J. Tolley, {\it {Complete Decoupling Limit of Ghost-free
  Massive Gravity}},  {\em JHEP} {\bf 1311} (2013) 059,
  [\href{http://arxiv.org/abs/1307.4769}{{\tt arXiv:1307.4769}}].

\bibitem{Hassan:2011zd}
S.~Hassan and R.~A. Rosen, {\it {Bimetric Gravity from Ghost-free Massive
  Gravity}},  {\em JHEP} {\bf 1202} (2012) 126,
  [\href{http://arxiv.org/abs/1109.3515}{{\tt arXiv:1109.3515}}].

\bibitem{DeFelice:2010pv}
A.~De~Felice and S.~Tsujikawa, {\it {Cosmology of a covariant Galileon field}},
   {\em Phys.Rev.Lett.} {\bf 105} (2010) 111301,
  [\href{http://arxiv.org/abs/1007.2700}{{\tt arXiv:1007.2700}}].

\bibitem{Khoury:2013tda}
J.~Khoury, {\it {Les Houches Lectures on Physics Beyond the Standard Model of
  Cosmology}},  \href{http://arxiv.org/abs/1312.2006}{{\tt arXiv:1312.2006}}.

\bibitem{Afshordi:2008rd}
N.~Afshordi, G.~Geshnizjani, and J.~Khoury, {\it {Do observations offer
  evidence for cosmological-scale extra dimensions?}},  {\em JCAP} {\bf 0908}
  (2009) 030, [\href{http://arxiv.org/abs/0812.2244}{{\tt arXiv:0812.2244}}].

\end{thebibliography}\endgroup

\appendix
\section{The Non-Relativistic Limit in an FRW Universe}
\label{sec:FRWnewt}

In this appendix we derive equations (\ref{eq:newttime}) and (\ref{eq:f5cos}) by expanding the Einstein and geodesic equations
around an FRW background. The FRW solution is sourced by the homogeneous energy momentum tensor
$T\mmm^\nm=\textrm{diag}(\rho_{{\rm
m}\,\infty},0,0,0)$\footnote{We could introduce other components apart from non-relativistic dust but this would not change the
conclusions. See the discussion in section \ref{sec:frwnewtlim}.} and the energy momentum tensor for the field $\phii$ given by
(\ref{eq:tmnphi}). The resultant equations are the Friedmann-scalar field system given by (\ref{eq:fried1})--(\ref{eq:fired3}). In
the Einstein frame, we have
\begin{equation}
 \dd s_{\rm E}^2=-\dd t^2+a(t)^2\dd\vec{x}^2
\end{equation}
whereas in the Jordan frame we have
\begin{equation}
 \dd s_{\rm J}^2=-A^2(\phii)\left(1-\frac{B\dot{\phi}\inff^2}{\Lambda^2}\right)\dd t^2+A^2(\phii)a(t)^2\dd\vec{x}^2.
\end{equation}
One can see that the effect of the disformal transformation is to add a lapse in the Jordan frame that cannot be absorbed into
the scale factor; the clocks for any particle species coupled to $\tg_\nm$ tick slower than those coupled to $g_\nm$ when the
disformal factor is different from zero. Setting $a_{\rm J}(t)=A(\phii)a(t)$ we have
\begin{equation}
 \tilde{\mathcal{N}}^2=\left(\frac{a_{\rm J}(t)}{a(t)}\right)^2\left(1-\frac{B^2\dot{\phi}\inff^2}{\Lambda^2}\right).
\end{equation}
Next, we want to add a non-relativistic, inhomogeneous source into the system. This is
described by  $T\mmm^\nm=\textrm{diag}(\rho_{{\rm m}\,\infty}+\rmm(r,t),0,0,0)$ and sources the perturbed metric
\begin{equation}
 \dd s^2_{\rm E}=\left[-1-2\Phi(r,t)\right]\dd t^2+a^2(t)\left[1-2\Psi(r,t)\right]\dd\vec{x}^2,
\end{equation}
in the Einstein frame and
\begin{equation}
 \dd s^2_{\rm J}=\left[-\tilde{\mathcal{N}}^2-2a_{\rm J}^2\Phi_{\rm J}(r,t)\right]\dd t^2+a_{\rm
J}^2(t)\left[1-2\Psi_{\rm J}(r,t)\right]\dd\vec{x}^2
\end{equation}
in the Jordan frame. We then wish to find the linearised (in metric potentials only) Einstein equations and the non-relativistic
geodesic equation in the Einstein frame. We begin by splitting the field into its homogeneous and inhomogeneous components i.e.
$\phi=\phii+\vpp$. The frame transformation law then gives
\begin{equation}
 \Phi_{\rm J}= \Phi+\alpha\vpp - \frac{B^2\dot{\phi}\inff\dot{\vpp}}{\Lambda^2}+\cdots\quad\textrm{and}\quad\Psi_{\rm
J}=\Psi-\alpha\vpp-\frac{B^2{\vpp^\prime}^2}{2a^2(t)\Lambda^2}+\cdots.
\end{equation}
We then have $BX/\Lambda^2\sim \pn^2$ and $\alpha\phi\sim\pn$, exactly as we found in section
\ref{sec:nrlim}. The only difference is that $X$ contains cross-terms between the homogeneous and inhomogeneous components of the
field due to the time-dependence of the background\footnote{Note also that $X=\dot{\phi}^2-(\nabla\phi)^2/2a^2$.}. The
perturbed (but not linearised) Einstein equations are:
\begin{align}
\nabla^2\Psi&=4\pi Ga^2\rho +3Ha^2\left(\dot{\Psi} +
H\Phi\right)\nonumber+\frac{a^2}{2}\partial_0\phii\partial^0\vpp+\frac{a^2}{4}\partial_0\vpp\partial^0\vpp\\&+\frac{a^2}{4}
\partial_i\vpp\partial^i\vpp+\frac{a^2}{2}\left[V(\phii+\vpp)-V(\phii)\right],\\
\partial_0\partial_i\Psi&=-H\partial_i\Phi-\frac{1}{2}\partial_0(\phii+\vpp)\partial_i\phi,\\
\left(\partial_i\partial_j-\frac{1}{3}\delta_{ij}
\nabla^2\right)\left(\Psi-\Phi\right)&=\partial_i\vpp\partial_j\vpp-\frac { 1 } { 3 }
\delta_{ij}\partial_k\vpp\partial^k\vpp\quad\textrm{and}\\
a^2\partial_0\partial_0\Psi+\frac{1}{3}\nabla^2\left(\Phi-\Psi\right)&=-\frac{a^2}{6}\left(\frac{3}{2}
\partial_0\vpp\partial_0\vpp+\frac{1}{2}\partial_k\vpp\partial^k\vpp-3\left[V(\phii+\vpp)-V(\phii)\right]
\right)\nonumber\\&-2a^2H\left(\dot { H } +3H\right)\Phi+a^2H^2\left(\dot {
\Phi}+3\dot{\Psi}\right).
\end{align}
As discussed by \cite{Hui:2009kc} appendix E, only the non-constant part of $V$ contributes to the perturbed equations and so we
have subtracted the background part, which sources the homogeneous component of the field. In the non-relativistic limit all
time-derivatives of the metric potentials vanish and we also have $\nabla^2\Psi\gg H^2\Phi$ on small scales. Demanding that all
scalar corrections to the equations are second order in the metric potentials we find exactly
equations (\ref{eq:phiords1})--(\ref{eq:phiord2}), in which case $\Psi=\Phi=\pn$. The scalar's equation of motion is then
precisely
\begin{equation}
 \nabla^2\vpp=8\pi Q(t)a^2(t)  G\rho\mmm,
\end{equation}
with $Q$ given by (\ref{eq:Qdef}). On time-scales far shorter than Hubble we can set $a(t)=1$. Finally, we need the
non-relativistic limit of the geodesic equation. Using the definition of $\mathcal{K}$ (\ref{eq:kdef}), the geodesic equation
becomes\footnote{Just as in the Minkowski space case, one finds that $\Gamma_{00}^0$ contains field-dependent quantities unless
one ignores time-derivatives, post-Newtonian terms and terms suppressed by factors of $H$.}
\begin{equation}
 \ddot{x}^i+2H\dot{x}^i=-\frac{1}{a^2}\nabla\pn-\frac{1}{a^2}Q\nabla\phi,
\end{equation}
where we have ignored time-derivatives of $\pn$. Setting $a(t)=1$ and assuming time-scales $\ll H_0$ so that we can ignore the
friction term we arrive at precisely (\ref{eq:f5cos}).

\section{Field Profile for Parallel Plates}
\label{sec:plates}

Here, we find the profile for two infinitely long parallel plates separated by a distance $d$. This is the configuration used by
many laboratory experiments such as the E\"{o}t-Wash experiment and Casimir force searches\footnote{In fact, the strongest
Casimir force constraints come from measuring the force between a plane and a sphere, but one can approximate the sphere as
another plate to a high degree of accuracy.}. We denote the
coordinate transverse to the plates as $z$ and place the first plate at $z=0$ and the second one at $z=d$. We take the density of
the plates to be constant and denote them using $\rho_1$ and $\rho_2$ respectively. In the interior
of the first plate we have
\begin{equation}\label{eq:plate1}
 \frac{\dd^2\vpp}{\dd z ^2}=\mu^2\vpp^2+8\pi GQ\rho_1.
\end{equation}
Ignoring solutions that diverge as $z\rightarrow-\infty$, the solution is
\begin{equation}\label{eq:sol1}
 \vpp(z)=De^{\mu z}-\frac{8\pi G Q\rho_1}{\mu^2}\quad z\le0,
\end{equation}
where $D$ is an integration constant. In the region between the two plates we take the density to be zero, in which case we have
\begin{equation}\label{eq:plate2}
 \frac{\dd^2\vpp}{\dd z ^2}=\mu^2\vpp^2
\end{equation}
and the solution is
\begin{equation}\label{eq:sol2}
 \vpp(z)=E e^{-\mu z} + F e^{\mu z}\quad 0\le z\le d,
\end{equation}
where $E$ and $F$ are integration constants. Finally, in the interior of the second plate we have
\begin{equation}\label{eq:plate3}
 \frac{\dd^2\vpp}{\dd z ^2}=\mu^2\vpp^2+8\pi GQ\rho_2
\end{equation}
so that the solution is
\begin{equation}\label{eq:sol3}
 \vpp(z)=He^{-\mu z}-\frac{8\pi G Q\rho_2}{\mu^2}\quad z\ge d,
\end{equation}
where $H$ is an integration constant and we have ignored the solution that diverges as $z\rightarrow\infty$. Matching both the
solutions and their derivatives at $z=0$ and $z=d$, the integration constants are:
\begin{align}
 D&=\frac{4  \pi G Q \rho _1}{\mu ^2}-\frac{4  G \pi  Q \rho _2}{\mu ^2}e^{-d \mu },\\
E&=-\frac{4  \pi  GQ \rho _1}{\mu ^2},\\
F&=-\frac{4  \pi  GQ \rho _2}{\mu ^2}e^{-d \mu }\quad \textrm{and}\\
H&=-\frac{4  \pi  GQ \rho _1}{\mu ^2}+\frac{4  G \pi  Q \rho _2}{\mu ^2}e^{d \mu }.
\end{align}
Using this, we can calculate the pressure (force per unit area) exerted by one plate on another. Consider the configuration where
we remove the plate at $z=d$. In this case the solution is identical to equation (\ref{eq:sol2}) with $F=0$. The term proportional
to $E$ is the field sourced by the plate and the term proportional to $F$ the correction due to the other plate. Therefore, the
force per unit mass on the plate at $z=0$ due to the plate at $z=d$ is $F_5=-Q\dd \vpp/\dd z|_{E=0}$. Laboratory experiments tend
to measure the pressure on the plates (force per unit area) and not the total force. This is then given by
\begin{equation}\label{eq:pressure_plates}
 P_\vpp=-\int_0^D\rho_1Q\frac{\dd\vpp}{\dd z}\approx \frac{4\pi G Q^2\rho_1\rho_2}{\mu^2}e^{-\mu d},
\end{equation}
where $D$ is the thickness of the plate. In practice, we expect $\mu d\ll1$ and one can ignore the exponential. In this case, the
total force is identical to that predicted by general relativity with $G\rightarrow G(1+2Q^2)$.

\end{document}